\documentclass[manuscript, screen]{acmart}
\usepackage{balance}
\usepackage{graphics}
\usepackage[T1]{fontenc}
\usepackage{times}
\usepackage{color}
\usepackage{booktabs}
\usepackage{textcomp}
\usepackage{xspace}
\usepackage{microtype}
\usepackage{ccicons}
\usepackage{todonotes}
\usepackage{amsthm}
\usepackage{amsmath}
\usepackage{subcaption}
\usepackage[normalem]{ulem}

\newcommand\halftab[1][.5cm]{\hspace*{#1}}
\newcommand\smallspace[1][.08cm]{\hspace*{#1}}

\newcommand\squid{\textsc{SQuID}\xspace}
\newcommand\sql{\textsc{SQL}\xspace}
\DeclareMathOperator{\argmax}{argmax}

\AtBeginDocument{%
  \providecommand\BibTeX{{%
    \normalfont B\kern-0.5em{\scshape i\kern-0.25em b}\kern-0.8em\TeX}}}

\makeatletter

\let\c@table\c@figure
\makeatother 

\setcopyright{acmcopyright}
\copyrightyear{2021}
\acmYear{2021}
\acmDOI{xx.xxxx/xxxxxxx.xxxxxxx}

\acmConference[CHI '21]{CHI '21: ACM SIGCHI}{May 8--13, 2021}{Yokohama, Japan}
\acmBooktitle{CHI '21: ACM SIGCHI, May 8--13, 2021, Yokohama, Japan}
\acmPrice{15.00}
\acmISBN{XXX-X-XXXX-XXXX-X/XX/XX}

\makeatletter
\def\@copyrightspace{\relax}
\makeatother
\makeatletter
\def\@mkbibcitation{\relax}
\makeatother

\begin{document}

% \title[A Comparative User Study on Query by Example vs. Structured Query Language]
% {Querying Relational Databases: A Comparative User Study on Query by Example vs. Structured Query Language}

\title[Example-Driven User Intent Discovery]
{Example-Driven User Intent Discovery:\\
Empowering Users to Cross the SQL Barrier Through Query by Example}

\author{Anna Fariha}
\email{afariha@cs.umass.edu}
\affiliation{%
  \institution{College of Information and Computer Sciences, University of Massachusetts, Amherst}
}

\author{Lucy Cousins}
\email{lcousins@cs.umass.edu}
\affiliation{%
  \institution{Department of Geosciences, University of Massachusetts, Amherst}
}

\author{Narges Mahyar}
\email{nmahyar@cs.umass.edu}
\affiliation{%
  \institution{College of Information and Computer Sciences, University of Massachusetts, Amherst}
}

\author{Alexandra Meliou}
\email{ameli@cs.umass.edu}
\affiliation{%
  \institution{College of Information and Computer Sciences, University of Massachusetts, Amherst}
}

\begin{abstract}
	Traditional data systems require specialized technical skills where users need
to understand the data organization and write precise queries to access data.
Therefore, novice users who lack technical expertise face hurdles in perusing
and analyzing data. Existing tools assist in formulating queries through
keyword search, query recommendation, and query auto-completion, but still
require some technical expertise. An alternative method for accessing data is
Query by Example (QBE), where users express their data exploration intent
simply by providing examples of their intended data. We study a
state-of-the-art QBE system called \squid, and contrast it with traditional
\sql querying. Our comparative user studies demonstrate that users with varying
expertise are significantly more effective and efficient with \squid than \sql.
We find that \squid eliminates the barriers in studying the database schema,
formalizing task semantics, and writing syntactically correct \sql queries, and
thus, substantially alleviates the need for technical expertise in data
exploration.

\end{abstract}




\maketitle

\pagestyle{plain}

\section{Introduction}
The proliferation of computational resources and data sharing platforms has
reached an ever-growing base of users without technical computing expertise,
who wish to peruse, analyze, and understand data. From astronomers and
scientists who need to analyze data to validate their hypotheses, all the way
to computational journalists who need to peruse datasets to validate claims and
support their reporting, the broad availability of data has the potential to
fundamentally impact the way domain experts conduct their work. Unfortunately,
while data is broadly available, data access is seldom unfettered. Existing
systems typically cater to users with sound technical computing and programming
skills, posing significant hurdles to technical novices, who do not have strong
technical background. \emph{Democratization} of computational systems demands
equal access to people of different skills and
backgrounds~\cite{DBLP:conf/vldb/Parameswaran20, pu2020program,
gulwani2017programming}.

\smallskip

\emph{User Scenario (Adapted from~\cite{ShenSIGMOD2014}).}
	 Consider a sales executive who needs to prepare a sales report over the
	 last week consisting of sales records indicating which customers bought which
	 products. Most enterprise databases are large and sales records are not
	 stored in a flat format (e.g., spreadsheet). Instead, such large-scale sales
	 information is usually split into multiple tables to achieve database
	 normalization, and stored within a database management system such as
	 PostgreSQL. Furthermore, the table contents are often encoded for compression
	 and reference purposes (e.g., product ID instead of product name). Therefore,
	 to generate the sales report, the sales executive will have to
	 (1)~familiarize themselves with the data organization (schema) to locate
	 relevant tables and understand the name encoding schemes, and (2)~pose a
	 query in the \sql language that is both syntactically and semantically
	 correct to obtain the desired sales records in the correct format (e.g.,
	 customer names and product names). These steps are challenging for the sales
	 executive who lacks a technical background, and thus, they would prefer to
	 bypass such complexities.
	 However, an enterprise information worker, such as this sales executive,
	 is often aware of a few \emph{examples} that should be present in the report.
	 They might remember that \texttt{John Smith} bought an \texttt{iPad} and
	 \texttt{Nora Shankar} bought a \texttt{Samsung smartphone} last week.
	 Certainly, they might not remember all sales records, but can an
	 \emph{example-based} interaction mechanism effectively assist this sales
	 executive in their task here, with just these examples? Furthermore, for
	 users with some technical skills, would such an interaction model still be
	 useful?
	 	
\smallskip

Example-based interactions have been explored as a method to bridge the
usability gap of computational systems that typically require precise programs
from users, such as in our user scenario above. Under the \emph{programming by
example} (PBE) paradigm (also known as \emph{programming by demonstration}),
instead of writing a precise program to specify their intent, users only need
to provide a few examples of the mechanism or result they
desire~\cite{cypher1995eager, lieberman2000programming, santolucito2019live,
DBLP:conf/cade/Gulwani16}. Prior work conducted user studies to contrast PBE
tools against traditional alternatives~\cite{DBLP:conf/chi/DrososBGDG20,
DBLP:conf/uist/MayerSGLMPSZG15, lee2017towards,
DBLP:conf/oopsla/SantolucitoGWP18}. However, none of them considered PBE tools
that are specifically designed for data exploration over relational databases.
We argue that \emph{query by example} (QBE), a facet of PBE focused on access
and exploration of relational data, has unique characteristics and poses
distinct challenges compared to general PBE methods. The focus of our work in
this paper is to study the effectiveness and usability of state-of-the-art QBE
against the traditional relational data access methods that rely on \sql
programs, through comparative user studies. We proceed to provide some
background on PBE and QBE systems, highlight the unique aspects of QBE that
have not been addressed by prior work and call for a targeted study, and
summarize our method and the contributions we make in this paper.

\smallskip

\noindent
\textbf{Programming by example (PBE): background and applications.}
The PBE paradigm is based on the intuitive premise that users who may lack or
have low technical skills, but have expertise in a particular domain, can more
easily express their computational desire by providing examples than by writing
programs under strict language specifications.
This is in contrast with traditional
program synthesis~\cite{gulwani2017program, jha2010oracle,
DBLP:conf/aaai/RazaG18}, which requires a high-level formal specification
(e.g., first-order logic) of the desired program. Example-driven program
synthesis has been effectively used for a variety of tasks, such as code
synthesis for data scientists~\cite{DBLP:conf/chi/DrososBGDG20}; data
wrangling~\cite{DBLP:series/natosec/Gulwani16},
integration~\cite{inala2017webrelate},
extraction~\cite{DBLP:conf/pldi/BarowyGHZ15, DBLP:conf/pldi/LeG14},
transformation~\cite{DBLP:conf/popl/Gulwani11, DBLP:conf/sigmod/HeGLWNCCZ18},
and filtering~\cite{DBLP:conf/oopsla/WangGS16}; data structure
transformation~\cite{feser2015synthesizing}; text
processing~\cite{DBLP:conf/uist/YessenovTMMGLK13},
normalization~\cite{DBLP:conf/ijcai/KiniG15}, and summarization~\cite{sudocu};
querying relational databases~\cite{ShenSIGMOD2014}, and so on.

\smallskip

\noindent \textbf{Query by example (QBE): the need for a new study.}
Example-driven interactions have also been explored in the context of
retrieving and exploring relational data, which led to the development of
\emph{query by example} (QBE) systems~\cite{DBLP:conf/sigmod/FarihaSM18,
DBLP:journals/pvldb/FarihaM19, ShenSIGMOD2014, PsallidasSIGMOD2015,
DeutchICDE2016}. In QBE systems, a user is expected to provide examples of the
data records they would like to retrieve, in place of providing a well-formed
query in the \sql language. The QBE system then infers the query the user
likely intended, and uses it to retrieve additional records from the database.
QBE is a special category of PBE that brings forth unique aspects and
challenges. We proceed to describe three significant distinctions that motivate
our comparative study evaluation of QBE systems.

First, the traditional mechanism for retrieving relational data requires not
only strong technical skills over the \sql language, but also familiarity with
the structural organization of the data, called a schema. Schemas can be very
complex, may contain domain-specific abstractions, differ from one database to
the next, and could also get modified over time. As a result, even expert users
with prior \sql experience can struggle to familiarize themselves with the
schema of a previously unseen dataset, leading to difficulties in data
exploration. Therefore, QBE needs to be studied from the perspective of users
with varied levels of expertise, and the study needs to investigate the pain
points specific to relational data access and exploration.

Second, the operational mechanisms in QBE systems fundamentally differ from
those in general PBE systems. Traditional PBE approaches often rely on
demonstration, where the mechanism to solve the intended task is demonstrated
by the user. In contrast, in QBE, the user gives examples of the intended
output and not the querying mechanism. Other PBE approaches rely on complete
input-output specifications: the user needs to provide, typically small, sample
inputs and outputs and the system infers their intended program. This mechanism
is also not possible in a data exploration setting, where the input data is
predetermined and typically large, and the user can only provide a small set of
examples of their intended query output. Since the set of examples in the QBE
setting is naturally incomplete, there is typically a much larger number of
queries (programs) that could be compatible with them, compared to the general
PBE setting; thus, the effectiveness of QBE systems needs to be explored with a
targeted study.

Third, the setting of data exploration has two characteristics that can have
significant impact in the performance of a QBE system: (1)~Since the user needs
to provide example records from the dataset at hand, domain expertise can have
a bigger impact in the user experience than in the general PBE setting.
(2)~Data exploration tasks can be vague and subjective, where a strict
specification is often hard or even impossible to derive even by experts; this
is a perspective not relevant to general PBE and not explored by prior studies.

\smallskip

\noindent \textbf{Our scope and method.} In this paper, we present findings
from our comparative user studies over a QBE tool and the traditional
\sql-based mechanism. For our study, we picked
\squid~\cite{DBLP:conf/sigmod/FarihaSM18, DBLP:journals/pvldb/FarihaM19} as the
QBE tool, since it offers the state-of-the-art QBE mechanism for exploring
relational databases. \squid is built on top of PostgreSQL, which is an
open-source relational database management system. Given a few examples of the
desired data, \squid discovers a \sql query by exploiting the semantic
similarities observed in the examples. Under the hood, \squid uses a
probabilistic model, which infers a query as the most likely explanation of the
provided examples. \squid and other QBE systems have broad applications in data
exploration~\cite{idreos2015overview}, query reverse
engineering~\cite{DBLP:journals/vldb/TranCP14}, and recommendation
systems~\cite{lu2015recommender}. 

We conducted two comparative user studies: (1)~a controlled experiment study
involving 35 participants, and (2)~an interview study involving 7 interviewees
to gain a richer understanding of users' issues and preferences. All
participants and interviewees had varying levels of \sql expertise and
experience, but were required to have at least basic \sql skills. Our studies
focused on the task of data exploration and explored how \squid compares
against the traditional \sql querying mechanism, over a variety of objective
and subjective data exploration tasks. Specifically, our study aimed to
identify the most critical issues users face when interacting with the
traditional \sql querying mechanism, to what extent a QBE system like \squid
can alleviate these challenges, how effective \squid is over a variety of data
exploration tasks, and what are the possible pain-points of \squid. 

\smallskip

\noindent \textbf{Contributions.} We summarize our contributions below:

\begin{itemize}
	 
	 \item Through an analysis of the \sql queries issued by the controlled
	 experiment study participants and quantitative analysis of the data collected
	 from the study, we found that participants were significantly more
	 \emph{effective} (achieved more accurate results) and \emph{efficient}
	 (required less time and fewer attempts) over a diverse set of subjective and
	 objective tasks using \squid compared to manual \sql programming.
	 
	 \item From observations made from the behavior of the interviewees during
	 our interview study, and their qualitative feedback, we identified three key
	 challenges that \sql poses to the users: familiarizing oneself with the
	 database schema, formally expressing the semantics of the task, and writing
	 syntactically correct queries. From the qualitative feedback of the
	 interviewees, we confirmed that \squid removes these \sql challenges
	 altogether and assists the users in effective data exploration. Notably, even
	 some of the \sql experts reported that certain subjective queries were
	 extremely hard to encode in \sql and that they would prefer \squid over \sql
	 in those circumstances.
	 
	 \item Finally, we discuss how \squid and traditional \sql mechanisms
	 complement each other, under what circumstances the users prefer one over the
	 other, and how the QBE tools should be expanded to achieve more user
	 acceptance. While our results validate some findings of prior
	 studies over other PBE approaches~\cite{DBLP:conf/oopsla/SantolucitoGWP18},
	 we contribute new empirical insights gained from our studies that indicate
	 that even a limited level of domain expertise (knowledge of a small subset of
	 the desired data) can substantially help overcome the lack of technical
	 expertise (knowledge of \sql and schema) in data exploration.
	
\end{itemize}

\noindent\textbf{Organization.} The rest of the paper is organized as follows:
We discuss the related work in Section~\ref{sec:related}.
Section~\ref{sec:squid-overview} gives an overview of the dataset and the two
systems used in our studies: \sql\footnote{\sql is a language that is the
querying mechanism standard of relational data management systems, but we
often, for ease of reference, refer to it as a system within the context of our
user studies.} and \squid. In Section~\ref{sec:design}, we describe the design
choices and methods of our comparative user studies. Section~\ref{sec:quan}
and~\ref{sec:qual} describe the quantitative findings and the qualitative
feedback found from the user studies, respectively. We discuss the key
take-aways from the user study and provide guidelines to improve QBE tools with
additional features in Section~\ref{sec:discussion}. Finally, we conclude in
Section~\ref{sec:summary}.

\section{Related Work}\label{sec:related}
In this section, we provide an overview of the existing PBE and QBE approaches,
discuss alternative mechanisms that also aid users in data exploration, and
discuss prior literature on comparative user studies over other PBE
approaches.

\subsection*{Programming-by-example (PBE) approaches} 
\looseness-1 Many PBE approaches have been developed in the literature to aid
novices or semi-experts in a variety of data management tasks. The focus of PBE
is to not only solve the task, but also provide the \emph{mechanism} that can
solve the task. To this end, all PBE tools learn from the user examples and
synthesize programs that can produce the desired results.
To help data scientists write complex data-wrangling and data-transformation
codes, WREX~\cite{DBLP:conf/chi/DrososBGDG20} proposes an example-driven
program synthesis approach. To enable integration of web data with
spreadsheets, WebRelate~\cite{inala2017webrelate} facilitates joining
semi-structured web data with relational data in spreadsheets using
input-output examples. FlashRelate~\cite{DBLP:conf/pldi/BarowyGHZ15} and
FlashExtract~\cite{DBLP:conf/pldi/LeG14} enable extraction of relational data
from semi-structured spreadsheets, text files, and web pages, using examples.
Data-transformation-by-example approaches~\cite{DBLP:conf/popl/Gulwani11,
DBLP:conf/sigmod/HeGLWNCCZ18} led to the development of the
FlashFill~\cite{FlashFill} feature in Microsoft Excel, which can learn the
user's data transformation intent only from a few examples. Beyond data
management tasks, recently, PBE has been used for text
processing~\cite{DBLP:conf/uist/YessenovTMMGLK13}, text
normalization~\cite{DBLP:conf/ijcai/KiniG15}, and personalized text
summarization~\cite{sudocu}. Live
programming~\cite{DBLP:conf/chi/SantolucitoHP19} helps novice programmers to
understand their codes, where they can manipulate the input by directly editing
the codes and manipulate the output by providing examples of the desired
output. Beyond computational tasks, PBE tools also support creative tasks such
as music creation by example~\cite{DBLP:conf/chi/FridGJ20}, where a software
takes a song as an example and allows the user to interactively mix the
AI-generated music.

\subsection*{Query by example (QBE), query reverse engineering (QRE), and
similar approaches}
Some QBE systems~\cite{ShenSIGMOD2014, PsallidasSIGMOD2015} focus on
identifying relevant relations and joins to compensate the user's lack of
schema understanding, but are limited to project-join queries. These systems
only exploit the structural similarities of the examples and ignore the
semantic similarities. QPlain~\cite{DeutchICDE2016} requires provenance of the
examples from the users to better learn the join paths. However, this requires
understanding of the schema, content, and domain of the data, which novice
users often lack.

Unlike QBE approaches that can work only with partial output (example), query
reverse engineering (QRE) approaches require the entire output with respect to
the original database. With this complete output specification, QRE can target
more expressive queries~\cite{ZhangASE2013, WangPLDI2017}, but only works for
very small databases and fails to scale to large databases. Some QRE approaches
require the user to specify a small input database and the corresponding
output, and constants in the query~\cite{WangPLDI2017}. However, this requires
complete schema knowledge and precise domain knowledge.
QRE~\cite{WeissPODS2017, barcelICDT2017, TranVLDB2014, FastQRE,
ZhangSIGMOD2013, TanVLDB2017, regalPlus2018, PanevEDBT2016} is less challenging
than QBE, because it is aware of the entire output, while typically only a
small fraction of the output is available for QBE. Thus, QRE systems can build
data classification models on denormalized tables~\cite{TranVLDB2014}, assuming
the user-provided examples as positive and the rest as negative. However, due
to lack of sufficient annotated data, similar techniques do not apply for QBE.

A problem similar to QBE in relational databases is set expansion in knowledge
bases~\cite{ZhangSIGIR2017, DBLP:conf/icdm/WangC07, wordgrabbag}.
SPARQLByE~\cite{SPARQLByE2016} allows querying datasets in resource description
framework (RDF) by annotated (positive/negative) examples. In semantic
knowledge graphs, systems exist to address the entity set expansion problem
using maximal-aspect-based entity model, semantic-feature-based graph query,
entity co-occurrence information, etc.~\cite{LimEDBT2013, JayaramTKDE2015,
HanICDE2016, MetzgerJIIS2017}. Although not applicable in the relational
domain, these approaches also exploit the semantic context of the examples;
however, they cannot learn new semantic properties that are not explicit in the
knowledge base.

\subsection*{Aiding novice users explore relational data}
Beyond by-example methods, alternative approaches exist to aid novice users
explore relational databases. Keyword-based search~\cite{AgrawalICDE2002,
HristidisVLDB2002, ZengEDBT2016} allows accessing relational data without
knowledge of the schema and \sql syntax, but does not facilitate search by
examples. Other notable systems that aim to assist novice users in data
exploration and complex query formulation are: QueRIE, a query recommendation
based on collaborative filtering~\cite{EirinakiTKDE2014}, SnipSuggest, a
context-aware \sql autocompletion system~\cite{KhoussainovaPVLDB2010},
SQL-Sugg, a keyword-based query suggestion system~\cite{FanICDE2011}, YmalDB, a
``you-may-also-like''-style data exploration system~\cite{YmalDBVLDB2013}, and
SnapToQuery, an exploratory query specification assistance
tool~\cite{JiangVLDB2015}. These approaches focus on assisting users in query
formulation, but assume that the users have sufficient knowledge about the
schema and the data. VIDA~\cite{DBLP:journals/debu/LeeP18},
ShapeSearch~\cite{DBLP:conf/sigmod/SiddiquiLWKP20}, and
Zenvisage~\cite{DBLP:journals/pvldb/SiddiquiKLKP16} are visual query systems
that allow visual data exploration, but they require the user to be aware of
the trend within the output. Some approaches exploit user interaction to assist
users in query formulation and result delivery~\cite{AbouziedPODS2013,
BonifatiTODS2016, DimitriadouTKDE2016, GeBigdata2016, LiVLDB2015}. There, the
user has to provide relevance feedback on system-generated tuples. However,
such highly-interactive approaches are not suitable for data exploration as
users often lack knowledge about the system-provided tuples, and thus, fail to
provide correct feedback reflecting their query intent. Moreover, such systems
often require a large number of user interactions.

\subsection*{User study of PBE approaches} 
\looseness-1 Drosos et al.~\cite{DBLP:conf/chi/DrososBGDG20} present a
comparative user study contrasting WREX against manual programming. The study
results indicate that data scientists are more effective and efficient at data
wrangling with WREX over manual programming. Mayer et
al.~\cite{DBLP:conf/uist/MayerSGLMPSZG15} presents comparative study between
two user interaction models---program navigation and conversational
clarification---that can help resolve the ambiguities in the examples in
by-example interaction models. Lee et al.~\cite{lee2017towards} presents an
online user study on how PBE systems help the users solve complex tasks. They
identify seven types of mistakes commonly made by the users while using PBE
systems, and also suggest an actionable feedback mechanism based on
unsuccessful examples. Santolucito et
al.~\cite{DBLP:conf/oopsla/SantolucitoGWP18} studied the impact of PBE on
real-world users over a tool for shell scripting by example. Their study
results indicate that while the users are quicker to solve the task using the
PBE tool, they trust the traditional approach more. However, none of these
studies focus on QBE in particular, which is a PBE system tailored towards data
exploration over relational databases. The performance of a QBE tool is
affected by additional factors, such as the subjectivity of the data
exploration task and the domain knowledge of the user. Moreover, traditional
data access and exploration methods pose hurdles not only to novices, but to
expert users as well. These factors indicate the need for a new study that
targets QBE systems in particular.

\section{Overview of the dataset and systems}\label{sec:squid-overview}
In our comparative user studies, we studied how users perceive a state-of-the-art QBE system,
\squid, compared to the traditional \sql querying mechanism, over a variety of
subjective and objective data exploration tasks. In this section, we provide an overview of the dataset
we used in our studies, along with brief description of both systems.

\subsection{Dataset} For our comparative user studies, our goal was to emulate
data exploration tasks in a controlled experiment setting. Generally, people
explore data they are interested in and within a domain they are somewhat
familiar with. Moreover, data exploration with QBE expects some basic domain
familiarity, as users need to be able to provide examples. Therefore, our goal
in selecting a dataset was to identify a domain of general interest, where most
study participants can be expected to have a basic level of domain familiarity.
Furthermore, the dataset needs to be sufficiently large to emulate the
practical challenges that users face during data exploration. We selected the
Internet Movie Database (IMDb)\footnote{IMDb: \url{www.imdb.com/}}, which
satisfies these goals. The IMDb website is well-known source of movie and
entertainment facts, has over 83 million registered users and about 927 million
yearly page visits.\footnote{IMDb.com Analytics:
\url{www.similarweb.com/website/imdb.com/}} The database contains information
regarding over 10 million personalities along with their demographic
information; and about 6 million movies and TV series, along with their genre,
language, country, certificate, production company, cast and crew, etc.

\subsection{Structured query language (SQL)}
\looseness-1 The traditional way to query a relational database is to write a
query in structured query language (\sql). \sql is one of the most widely-used
programming languages (54.7\% developers use \sql~\cite{liu_2020}) for handling
structured data, is specifically designed to query relational databases,
and has been used for over 50 years. \sql is a declarative query language and is
primarily based on relational algebra. The \sql language consists of several
elements such as clauses, expressions, predicates, statements, integrity
constraints, etc. \sql has been implemented by different developers---such as
Oracle, Microsoft SQL, MySQL, PostgreSQL, etc.---slightly differently, however,
fundamentally, they all work the same way. For our comparative user studies, we
picked PostgreSQL, which is a free and open-source relational database
management system.

\begin{figure}[t!]
	\centering
	\includegraphics[width=1\linewidth]{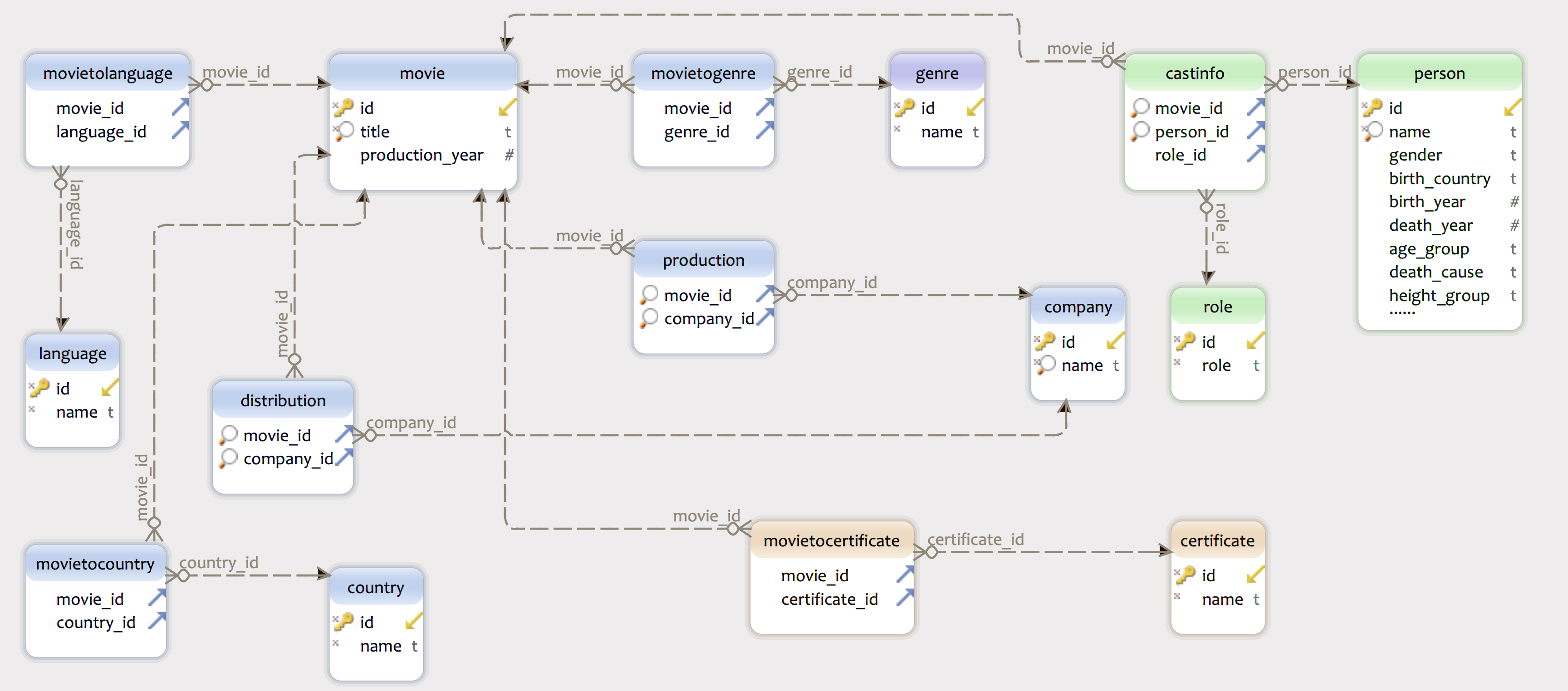}	
		 \vspace{-5mm}
	 \caption{Complete schema of the IMDb database with 8 main relations:
	 \texttt{movie}, \texttt{person}, \texttt{genre}, \texttt{language}, \texttt{country}, \texttt{company}, \texttt{role}, and \texttt{certificate}; 
	 and 7 connecting relations that associate the main relations: \texttt{castinfo}, \texttt{movietogenre}, \texttt{movietolanguage}, 
	 \texttt{distribution}, \texttt{movietocountry}, \texttt{movietocertificate}, and \texttt{production}.}
	\label{fig:schema}
\end{figure}

Relational databases usually organize data in a \emph{normalized} form, to
avoid redundancy. This is in contrast with the flat data format where all
attributes of an entity are stored together within the same row. For example,
the detailed schema of the IMDb database, split in 15 relational tables, is
shown in Figure~\ref{fig:schema}. Here, the relation \texttt{movie} contains
only three attributes about movies: a numerical record \texttt{id} (called
\emph{primary key}), a text attribute specifying the \texttt{title} of the
movie, and the \texttt{production year} of the movie. However, information
about associated genres of a movie is not present in the \texttt{movie} table.
To figure out the genres of a movie, one would need to write a \sql query to
\texttt{JOIN} the tables \texttt{movie}, \texttt{movietogenre}, and
\texttt{genre}. The query would also need to specify the \emph{logic} behind
this join, i.e., which rows in the \texttt{genre} table are relevant to a
particular movie in the \texttt{movie} table.
\looseness-1 \sql is a relatively simple language with a limited set of operators (e.g.,
\texttt{SELECT}, \texttt{PROJECT}, \texttt{JOIN}, etc.). While this simplicity
enables the users to learn quickly how to express easy intents using \sql
(e.g., the \sql query \texttt{SELECT title FROM movie} would retrieve all movie
titles), it comes at the cost that complex intents are hard to express in \sql.
Specifically, the restrictions in the data organization (normalized schema) and
the simplicity of the \sql operators make complex tasks harder to translate in
\sql: it requires the users to specify the entire data-retrieval logic.
Overall, writing a successful \sql query for a data exploration tasks requires
several skills: (1)~familiarity with the database schema, (2)~understanding
of the table semantics, (3)~understanding of the \sql operators, (4)~knowledge
of the \sql syntax, and (5)~expertise in translating task intents to \sql.

\subsection{SQuID}\label{squid}

\squid~\cite{DBLP:conf/sigmod/FarihaSM18, DBLP:journals/pvldb/FarihaM19} is an
end-to-end system that automatically formulates complex \sql queries over
commonly used operators and functions---such as \texttt{SELECT}, \texttt{FROM},
\texttt{WHERE}, \texttt{JOIN}, \texttt{GROUP BY}, \texttt{INTERSECT},
\texttt{HAVING}, \texttt{COUNT}, etc.---based on a few user-provided examples.
\squid does not require the users to have any knowledge of the database schema
or the query language. The key mechanism of \squid is to extract the
\emph{semantic similarity} of the example tuples (e.g, all example entities are
Male actors), express them in terms of \emph{selection predicates} (e.g.,
\texttt{Gender = Male}), and then construct a \sql query that includes an
appropriate subset of those selection predicates. To figure out the appropriate
subset of selection predicates, \squid distinguishes \emph{coincidental}
properties from the intended ones. Intuitively, if a property observed in the
example entities is very common over the entities of the database, then it is
unlikely to be intended and more likely to be coincidental. For example, if
90\% of the people in a database have black hair and the user provides 3
examples where all of them have black hair as well, \squid assumes that this is
just a coincidence and not a genuine intent. In contrast, if a property
observed in the example entities is rarely observed over the entities in the
database, then it is more likely to be intended. For example, if only 5\% of
the people in a database have green eyes and all the user examples also have
green eyes, \squid interprets it as a genuine user intent.

\looseness-1 \squid expresses the problem of query intent discovery using a
\emph{probabilistic model} that infers the most likely query, given the
examples. To mathematically derive the intended query, \squid applies
\emph{abduction}~\cite{abductionMenzies96, DBLP:reference/ml/Kakas17}, an
inference method that aims to find the most likely explanation (query intent)
from an incomplete observation (examples). Unlike deduction, the premises do
not guarantee the conclusion in abduction. A deductive approach would report
all queries whose results contain the examples. While it guarantees that the
user's intended query definitely resides within the reported queries, such an
approach is of no practical use when the number of reported queries is large.
In contrast, thanks to abduction, \squid finds the most likely query intent,
given the examples. Formally, given a database ${D}$ and a set of examples $E$,
\squid returns the query $Q = \argmax_q Pr (q \mid E)$ such that $E \subseteq
Q(D)$, where $Q(D)$ denotes the set of tuples in the result of $Q$ over $D$,
and $Pr(q \mid E)$ is the probability of $q$ to be the intended query, given
the example set $E$.

\begin{figure}[t!]
	\centering
	\vspace{-5mm}
	\includegraphics[width=0.6\linewidth]{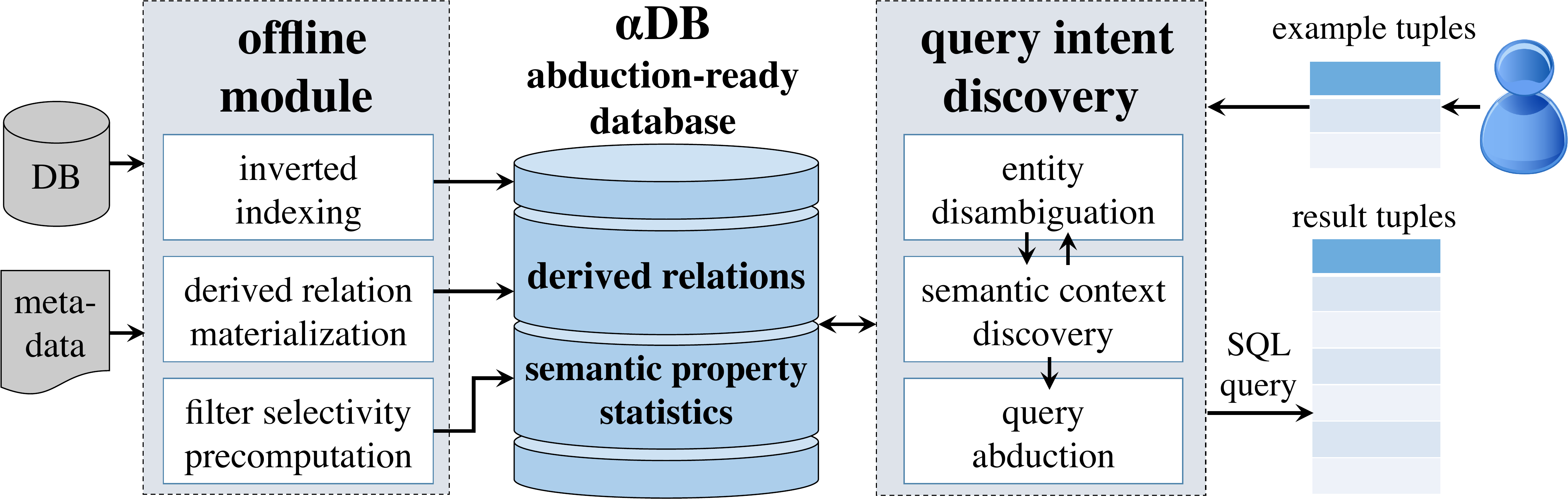}	
	
	 \caption{(Adapted from~\cite{DBLP:journals/pvldb/FarihaM19}) \squid's
	 offline module constructs an abduction-ready database that stores derived
	 relations and semantic property statistics. The query intent discovery module
	 serves the user: it takes the user-provided examples, consults the
	 abduction-ready database, discovers the most likely query, executes it, and
	 returns the results to the user.}
	\label{fig:archi}
	\vspace{-5mm}
\end{figure}

Figure~\ref{fig:archi} depicts \squid's system architecture. To achieve
real-time performance, \squid relies on an offline precomputation strategy that
stores semantic properties of all entities of the database and the
corresponding statistics of those semantic properties (e.g., how frequently a
semantic property is observed in the database) in an \emph{abduction-ready}
database. During the online query intent discovery phase, \squid consults the
abduction-ready database to derive relevant semantic properties based on the
provided examples, and applies abduction to select the optimal set of
properties towards constructing the most likely query. Finally, \squid executes
the inferred query and presents the results to the user.

\smallskip

\emph{Example User Scenario (Adapted from~\cite{DBLP:journals/pvldb/FarihaM19}).}
	 \looseness-1 A user provides the example set \{\texttt{Robin Williams},
	 \texttt{Jim Carrey}, \texttt{Eddie Murphy}\} to query the IMDb database using
	 \squid in search for ``funny'' actors (Figure~\ref{fig:squidui}). \squid
	 discovers the following semantic similarities among the examples: (1) all are
	 \texttt{Male}, (2) all are \texttt{North American}, and (3) all appeared in
	 more than 40 \texttt{Comedy} movies. Among these, \texttt{Male} and
	 \texttt{North American} are very common in the database. In contrast, a very
	 small fraction of actors in the database are associated with such a high
	 number of \texttt{Comedy} movies; this means that it is unlikely for this
	 similarity to be coincidental, as opposed to the other two. Based on
	 abduction, \squid selects the third similarity as the best explanation of the
	 observed example tuples, and produces the following \sql query:\\
{\small
\noindent \smallspace \phantom{\texttt{Q4:$\;\;$}} \noindent \texttt{SELECT person.name\\
\noindent \smallspace \smallspace \halftab \smallspace \smallspace FROM person, castinfo, movietogenre, genre\\
\noindent \smallspace \smallspace \halftab \smallspace \smallspace WHERE person.id = castinfo.person\_id\\
\noindent \smallspace \smallspace \halftab \halftab AND castinfo.movie\_id = movietogenre.movie\_id\\
\noindent \smallspace \smallspace \halftab \halftab AND movietogenre.genre\_id = genre.id\\
\noindent \smallspace \smallspace \halftab \halftab AND genre.name = `Comedy'\\
\noindent \smallspace \smallspace \halftab \smallspace \smallspace GROUP BY person.id\\
\noindent \smallspace \smallspace \halftab \smallspace \smallspace HAVING COUNT(*) >= 40}
}

\squid then executes this query and presents the results containing two
well-known funny actors---\texttt{Adam Sandler} and \texttt{Ben
Stiller}---among others (Figure~\ref{fig:squidui}).

\section{Evaluation: Comparative User Study}\label{sec:design}
\begin{figure}[t!]
	\centering
	\includegraphics[width=1\linewidth]{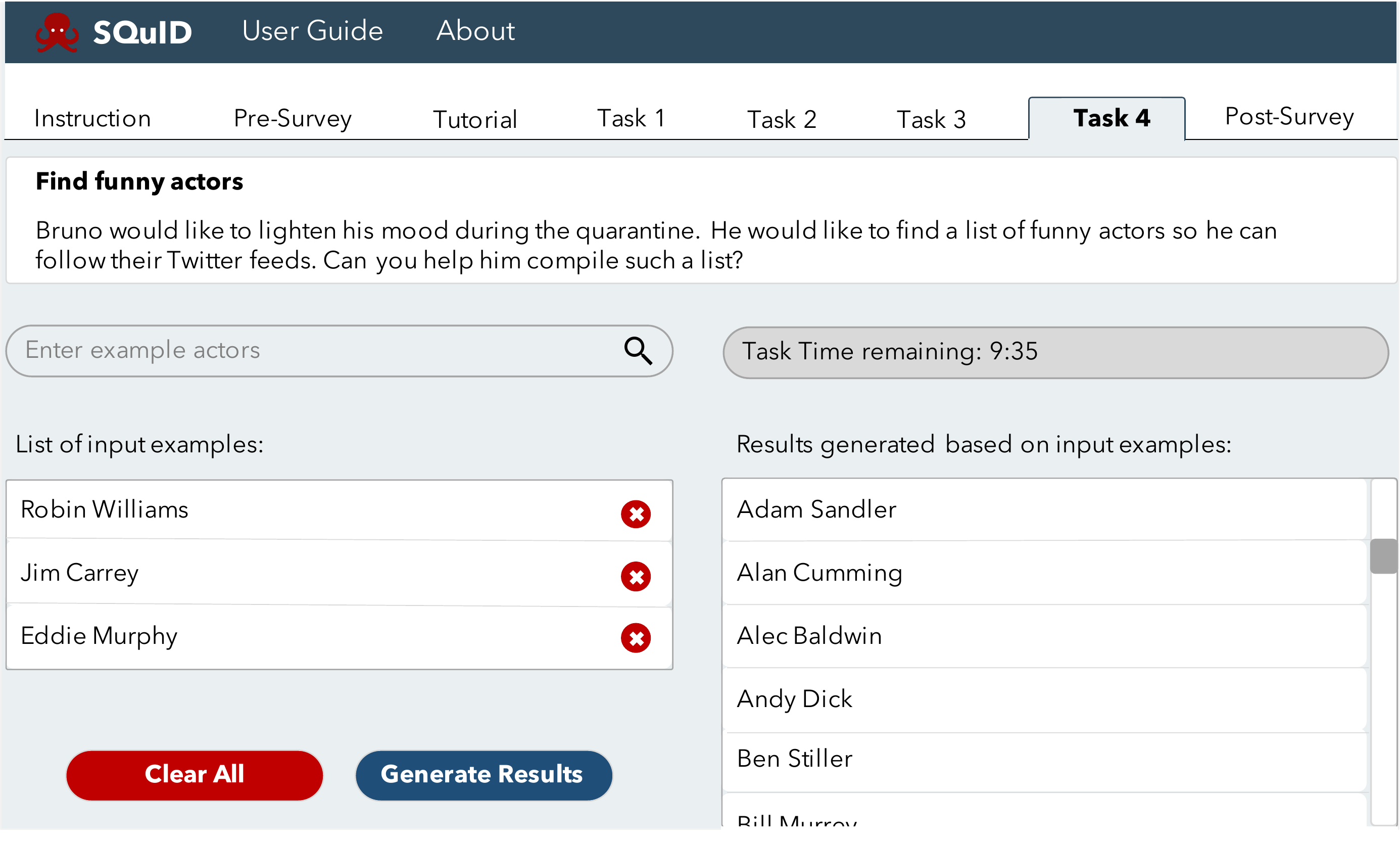}	
		 \vspace{-6mm}
	 \caption{The graphical user interface of \squid used in our user study.
	 The task description is at the top. The left panel allows the users to
	 provide examples with an auto-completion feature. 
	 \squid infers the user's intended query from the examples,
	 executes it, and shows the results in the right panel. The result set
	 contains more actors, but we only show the first five (alphabetically) here.}
	 \vspace{-2mm}
	\label{fig:squidui}
\end{figure}

\looseness-1 In our user studies, our goal was to quantitatively compare the
efficacy and efficiency of \squid and \sql over a variety of data exploration tasks,  while also
gathering qualitative feedback from users regarding their experiences with the
systems. To this end, we opted for two separate comparative user studies: (1)~a
controlled experiment study, with a fixed set of tasks, over a
group of participants of sufficient size to support quantitative evaluation; (2)~an interview
study, with a flexible set of tasks, over a small group to gather qualitative
user feedback. Due to the situation caused by the current COVID-19 pandemic,
both studies were conducted online: the controlled experiment was conducted
through a website, hosted on our university servers, and the interview study was
conducted over Zoom.

For both studies, we provided the database schema (Figure~\ref{fig:schema}) and
a graphical user interface with a text box, where the participants could write
\sql queries to interact with a PostgreSQL database. For \squid, we provided a
graphical user interface to allow the participants to interact with the system
(Figure~\ref{fig:squidui}). We now proceed to describe the settings, design
choices, and methods of our comparative user studies. We first describe our
controlled experiment study over a user group of 35 participants, followed by
our interview study with a smaller group of 7 interviewees.

\subsection{Study 1: controlled experiment study}\label{gt}

\subsubsection*{Participants} For our controlled experiment study, we recruited
students who were enrolled in an undergraduate computer science course on Data
Management Systems at our university during the Spring 2020 semester. The
course offers an introduction to data management systems and the \sql language.
This ensured that our study participants would have basic familiarity with
\sql, which is required to compare the two systems: \squid and \sql. We invited
all 89 students enrolled in this course to take part in the study and 35 of
them agreed to participate. We offered extra credit for study participation;
students who opted to not participate were given alternative opportunities for
extra credit. We labeled these participants P1--P35. The average grade the
participants achieved in the course was 86.3 (out of 100), with a minimum grade
of 45, and a maximum grade of 100; the standard deviation of the grades was
9.87. This indicates a broad range in our participants' \sql skills, which was
one of our goals. While all of them had prior experience and exposure, some had
only very basic skills (and failed the class) and some achieved advanced skills.

\subsubsection*{Tasks} \looseness-1 We designed 4 data exploration tasks over
the IMDb database. Our goal was to observe what challenges a set of diverse
tasks poses to the participants and how the challenges vary based on the
subjectivity of the tasks and the mechanism (\squid or \sql) used to solve the
tasks. To this end, we designed two objective tasks: (1)~to find \emph{Disney}
movies and (2)~to find \emph{Marvel} movies; and two subjective tasks: (1)~to
find \emph{funny} actors and (2)~to find \emph{strong} and muscular actors. We
provided a detailed description for each task to the study participants.
(Details are in our supplementary materials.)

\subsubsection*{Task-assignment mechanism} \looseness-1 Each participant was
assigned all of the four tasks in the sequence: Disney, Marvel, funny, and
strong. This order was enforced to ensure that they perform objective tasks
first, which are easier, and then move to more complex and subjective tasks. We
randomized task-system pairings to make sure that for each task, about half of
the participants use \squid while the other half use \sql. The task-assignment
mechanism was as follows: ~for each user, we randomized which system (\squid or
\sql) they are allowed to use for each task. Everyone did the tasks---Disney,
Marvel, funny, strong---in that order, but there were two possible system
assignment orders: (a)~SQuID, SQL, SQuID, SQL, or (b)~SQL, SQuID, SQL, SQuID.
Each participant was randomly given one of these assignments. This resulted in
randomized task-system pairings, with the constraint that each participant must
solve one objective and one subjective task using \sql and the remaining two
tasks (also one objective and one subjective) using \squid. This mechanism also
eliminated any potential order bias with respect to the treatment system as
half of the participants interacted with \squid before \sql, while the other
half interacted with \sql before \squid. Within each task (e.g., Disney), each
participant used either \squid or \sql to solve each task, but not both.

\subsubsection*{Study procedure} \looseness-1 This study was conducted online
and the participants took the study over the Internet on a specific website,
hosted on our university servers. We sent out the URL of the website during
recruitment. At the beginning of the study, participants were asked a series of
questions about their familiarity with \sql. The questions asked the
participants to provide answers using a 5-point Likert-scale ranging from ``Not
familiar (1)'' to ``Very familiar (5)''. Next, there was a question asking them
at what frequency they watch movies, followed by a questions about overall
movie and actor familiarity where participants could select multiple options.
After this survey, participants were given an interactive tutorial, which was
divided into two sections, walking them through the steps to obtain results
with both \squid and \sql. The tutorial took about 2--5 minutes to complete.
After the tutorial, the participants started the tasks. They had 10 minutes for
each task, but could finish before the time was up if they chose to.
Participants were asked to avoid using Internet search, but if they did, they
were encouraged to report it. After each task, the participants were asked to
answer a post-task survey with two questions: the first one was about the
difficulty of the task where the participants had to provide answers using a
5-point Likert-scale ranging from ``Very difficult (1)'' to ``Very easy (5)'';
and the second one was about their satisfaction with the results where the
participants had to provide answers using a 5-point Likert-scale ranging from
``Very unsatisfied (1)'' to ``Very satisfied (5)''. After completing all four
tasks, the participants were asked to answer four survey questions: the first
one was regarding their preferences between \squid and \sql where the
participants had to provide answers using a 5-point Likert-scale ranging from
``Definitely \sql (1)'' to ``Definitely \squid (5)''; the second one was about
usability comparison between \sql and \squid where the participants had to
provide answers using a 5-point Likert-scale ranging from ``\sql was a lot
easier (1)'' to ``\squid was a lot easier (5)''; the third one was about
satisfaction with results obtained using \squid where the participants had to
provide answers using a 5-point Likert-scale ranging from ``very unsatisfied
(1)'' to ``very satisfied (5)''; and the fourth one was about accuracy of the
results obtained using \sql where the participants had to provide answers using
a 5-point Likert-scale ranging from ``very inaccurate (1)'' to ``very accurate
(5)''.

\subsubsection*{Data collection} During the study, we collected all survey
responses and all inputs the participants provided to the systems.
Specifically, for \sql, we collected all their queries, including any
intermediate queries that they used to reach their final query; for \squid, we
collected all the examples they provided, along with the revision history
(addition or removal of examples). We stored all this information in JSON
format.

\subsubsection*{Data analysis} \looseness-1 During our data analysis, we
extracted the JSON data programmatically through Python scripts and implemented
custom functions to programmatically analyze the data. To quantitatively
evaluate the tasks performed by the participants, we compared their results
against the ground-truth results. We collected the ground-truth data from
publicly available lists on the IMDb website. For the objective tasks (Marvel
and Disney), we determined the ground truth by selecting one list for each. For
the subjective tasks (funny and strong), we compiled a list by combining seven
different lists for each. We selected lists that meet the following criteria:
(1)~they have a number of entries that is representative of the task (e.g.,
there are more than five Marvel movies, thus the list should contain more than
five entries), (2)~they are frequently-viewed, and (3)~they contain entries
that match the task objectives. For instance, we collected a list of 300 funny
actors, which was compiled from 7 shorter lists of funny actors. One of these
lists, titled ``Funny Actors'', has over 400,000 views, and includes 60
well-known comedians including Jim Carrey, Robin Williams, Eddie Murphy, Mel
Brooks, and Will Ferrell.\footnote{Funny Actors:
\url{https://www.imdb.com/list/ls000025701}} We provide all the lists we used
in our supplementary materials.

\begin{figure}
	\centering
	{\small
	\begin{tabular}{clllll}
		\toprule
		Interviewee ID & Gender  & Country of origin &  Program level  & \sql expertise & Area of specialization \\
		\midrule                          
		I1             & Female  & Greece            & 2nd year PhD    &  Medium 	   & Data management              		\\
		I2             & Male    & India             & 3rd year PhD    &  Low		   & Natural language processing        \\
		I3             & Male    & Hong Kong         & 2nd year MS     &  High		   & Systems       						\\
		I4             & Female  & China             & 5th year PhD    &  High		   & Data privacy       	\\
		I5             & Male    & India             & 4th year PhD    &  High		   & Theory and data management			\\
		I6             & Female  & Japan             & 2nd year PhD    &  Medium	   & Data privacy         \\
		I7             & Male    & USA               & 4th year PhD    &  High		   & Data privacy         \\
		\bottomrule
	\end{tabular}
	}
	\vspace{-2mm}
\caption{Demographic and experience details of the interviewees who participated in our interview study.}
\label{fig:intervieweedetails}
\end{figure}

\subsection{Study 2: interview study} We conducted a comparative interview
study to gain richer insights on users' behavior, their preferences, and issues
they faced while solving the data exploration tasks using both systems.

\subsubsection*{Interviewees} 
We recruited 7 interviewees for this study by targeting a diverse set of
computer science graduate students directly working or collaborating with the
data management research lab at our university. Out
of the 7 interviewees, 4 were male and 3 were
female; 6 of them were international students; and their ages ranged from 25 to
30 years old. All of them had experience using \sql for at least one year,
however, their expertise varied from moderate to expert. We label the
interviewees I1--I7. We provide further details on the interviewees in
Figure~\ref{fig:intervieweedetails}.

\subsubsection*{Tasks} \looseness-1 For this study, we asked the interviewees
to pick one objective task from the following list: (1)~Disney movies,
(2)~Marvel movies, (3)~animation movies, (4)~sci-fi movies, (5)~action movies,
(6)~movies by an actor of their own choice, or (7)~movies by a country of their
own choice. We also asked them to select one subjective task form this list:
(1)~funny actors, (2)~physically strong actors, or (3)~serious actors. The
variety of tasks allowed interviewees to pick tasks based on their interests
and enabled us to observe how the two systems compare over a variety of data
exploration tasks. This study was within-subject, i.e., all of the interviewees
were required to use both the systems (\squid and \sql) to solve each task.

\subsubsection*{Study procedure} \looseness-1 For each interview, two of our
research team members were present, one as primary to lead the interview and
ask questions and another as secondary to take notes and ask potential
follow-up questions. At the beginning of the study, we provided them the URL of
the study website over the chat feature of Zoom. During the study, the
interviewees first completed an interactive tutorial and then they were asked
to pick two tasks. The interviewees were then asked to solve each task using
both \squid and \sql, so that they can directly contrast the two systems. We
asked them to complete each task first using \squid and then using \sql, so
that the examples they would provide while using \squid would be free from
biases due to observing the results from their \sql query outputs. We did not
expose through the \squid interface the query that \squid generates, thus
avoiding biases when the interviewees were completing the \sql tasks. The
interviewees followed a think-aloud protocol and shared their screen over Zoom
during the study. They were observed by two interviewers who also asked
open-ended questions to the interviewees on completion of each of the two tasks
using both systems. The questions aimed to gather information on which of the
two systems the interviewees prefer, under what circumstances they prefer one
over the other, and the justification of why they do so. They were also asked
what challenges they faced while using the systems and whether some particular
task exacerbated these challenges. Finally, they were asked what type of
results they prefer during data exploration: specific or general.

\subsubsection*{Data collection} We recorded all interview sessions. The 7
interviews summed to 467 minutes. On average, each interview lasted about 67
minutes, with the shortest interview lasting 43 minutes and the longest one
lasting 77 minutes. Upon completion, we replayed the interview recordings,
manually transcribed the responses, and stored them as plain text in a
spreadsheet, resulting in 119 responses in total.

\subsubsection*{Data analysis} We thematically analyzed the responses using our
coding software (spreadsheet). Two independent coders from our team
independently coded the data. The following six themes emerged after several
rounds of analysis:
(1)~struggle in task understanding,
(2)~struggle in familiarizing oneself with the schema while using \sql,
(3)~difficulties with writing syntactically correct \sql queries,
(4)~struggle with solving vague/subjective tasks using \sql,  
(5)~struggle due to lack of domain familiarity while using \squid, and
(6)~preference between precision and recall of the results.
Inter-coder reliability was 0.98, calculated using Krippendorff’s alpha.

\section{Quantitative Results from Controlled Experiment}\label{sec:quan}
In this section, we present the quantitative results of the controlled
experiment study, summarizing our findings.

\subsection*{Participants had basic domain knowledge and SQL familiarity}
\begin{figure}[t]
	\centering
	\begin{subfigure}{.5\textwidth}
			\centering
			\includegraphics[height=38mm]{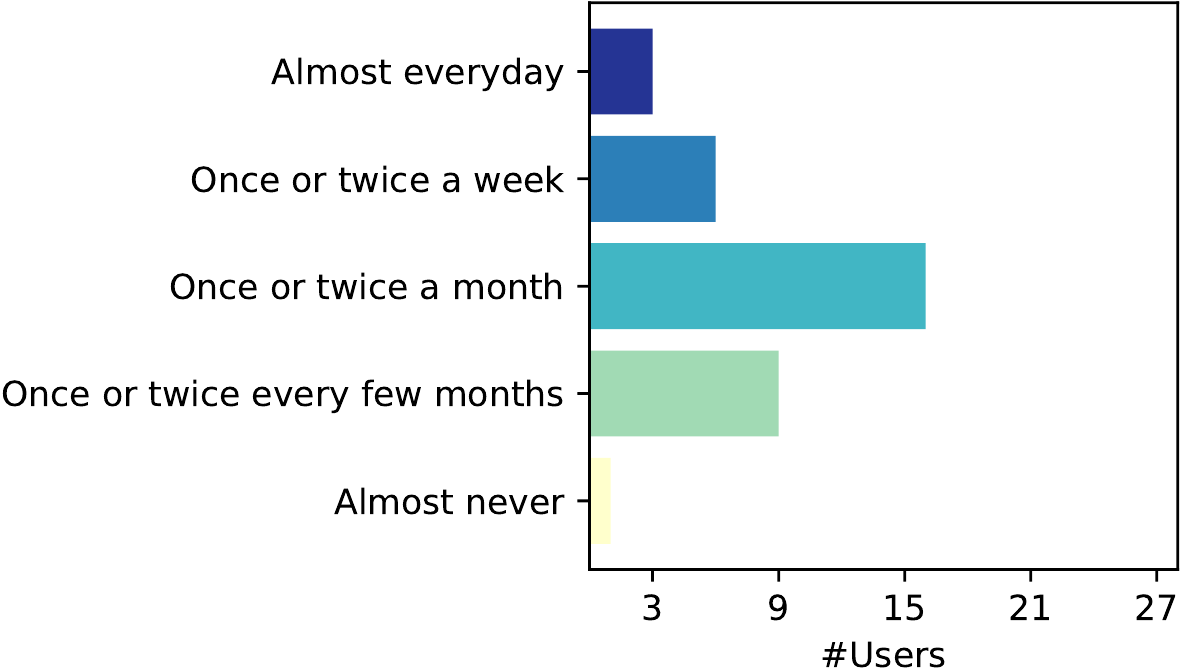}		
			\caption{Movie-watching frequency.}
			\label{fig:movieFrequency}
	\end{subfigure}	
	\hspace{3mm}	
	\begin{subfigure}{.45\textwidth}
			\centering
			\includegraphics[height=38mm]{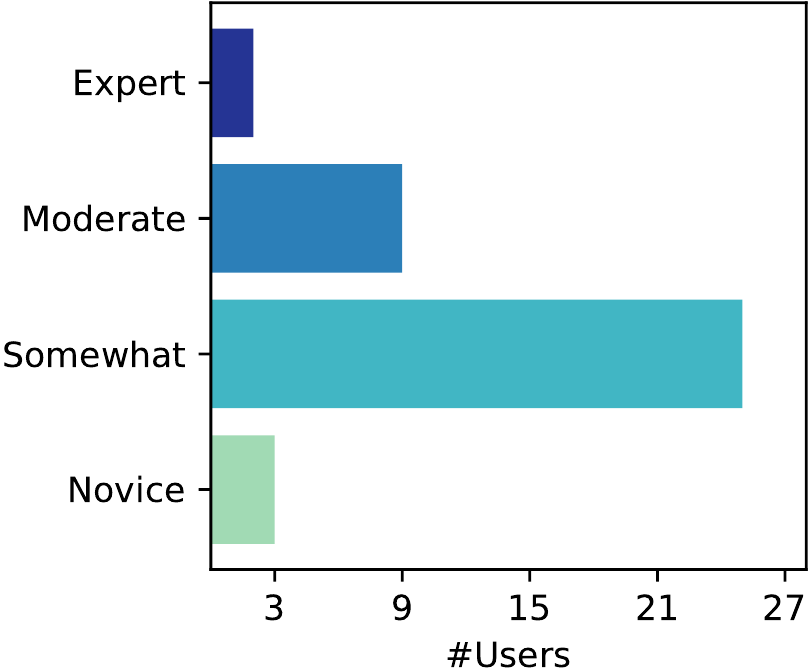}		
			 \caption{Overall knowledge of movies and actors.}	 
			 \label{fig:movieknowledge}
	\end{subfigure}
	\caption{Domain knowledge of the participants.}
	\vspace{-4mm}
\end{figure}

\looseness-1 The distribution of self-reported movie-watching frequency among
the participants is shown in Figure~\ref{fig:movieFrequency}, with the most
common response being `once or twice a month', followed by `once or twice every
few months'. The responses regarding actor and movie familiarity are summarized
in Figure~\ref{fig:movieknowledge}: a vast majority of the participants (25 out
of 35) reported that they were `somewhat' familiar with movies and actors. This
validates our choice of the IMDb database for conducting the study, as indeed,
we observed sufficient domain knowledge among the participants. Regarding \sql
expertise, all 35 participants reported being very familiar with easy \sql
queries and 34 reported being very familiar with moderately-complex \sql
queries. When asked regarding familiarity with complex \sql queries, 27
participants reported being very familiar, 6 were unsure, and 2 were not
familiar.

\subsection*{SQuID is generally more effective than \sql in generating accurate
results} \looseness-1 To quantitatively measure the quality of the results
produced by both \squid and \sql, we checked them against the ground-truth
results (discussed in Section~\ref{gt}). We used three widely-used correctness
metrics to quantify the result quality: precision, recall, and F1 score. These
metrics capture different aspects: precision captures ``preciseness'', i.e.,
the fraction of retrieved tuples that are relevant; recall captures
``coverage'', i.e., the fraction of relevant tuples that are correctly
retrieved; and F1 score---which is a harmonic mean of precision and
recall---maintains a balance between them.

\begin{figure}[t]
	\centering
	\begin{subfigure}{.31\textwidth}
		\centering
		\includegraphics[width=1\linewidth]{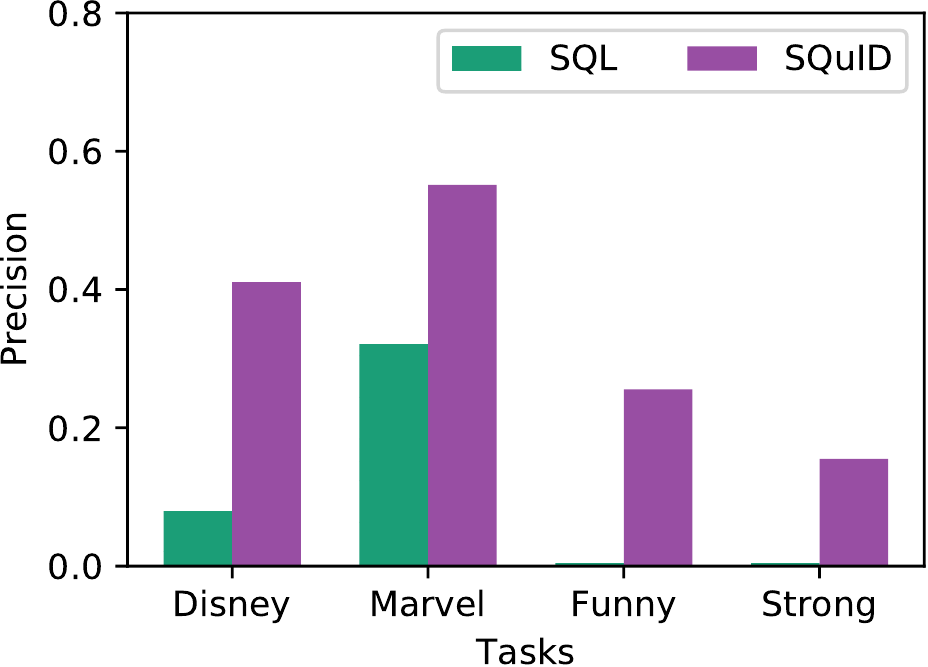}		
		\caption{\squid achieves higher precision than \sql for all tasks.}
		\label{fig:precision}
	\end{subfigure}
	\hspace{3mm}
	\begin{subfigure}{.31\textwidth}
		\centering
		\includegraphics[width=1\linewidth]{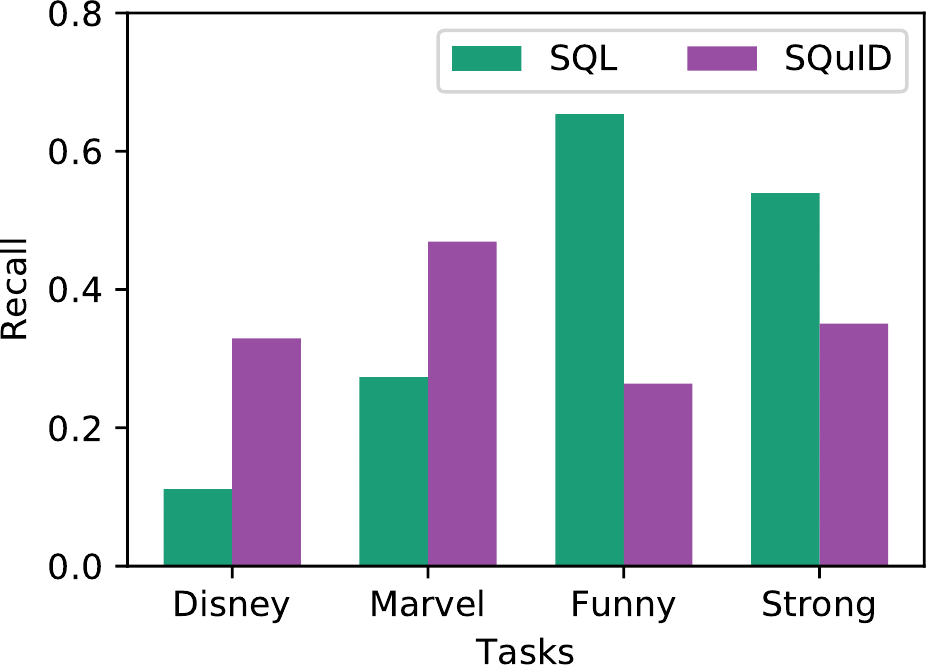}		
		 \caption{\squid achieves higher recall than \sql in two tasks.}	 
		\label{fig:recall}
	\end{subfigure}
	\hspace{3mm}
	\begin{subfigure}{.31\textwidth}
		\centering
		\includegraphics[width=1\linewidth]{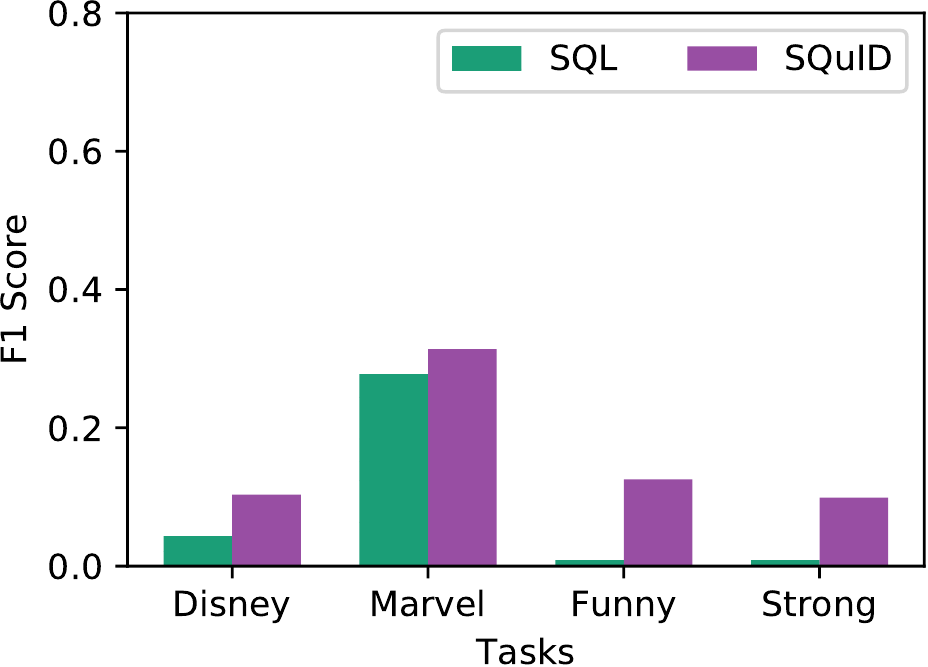}		
		 \caption{\squid achieves higher F1 score than \sql for all tasks.}	 
		 \label{fig:fscore}
	\end{subfigure}	
	\caption{Comparison of \squid vs. \sql in terms of average precision, recall, and F1 score.}
	\label{fig:accuracy}
	\vspace{2mm}
	\centering
	\begin{tabular}{l|ll|ll|ll}
		\toprule
		\multicolumn{1}{c}{} & \multicolumn{2}{c}{Precision} & \multicolumn{2}{c}{Recall} & \multicolumn{2}{c}{F1 Score}\\
		\midrule
		 \multicolumn{1}{c}{Task} & p-value & \multicolumn{1}{c}{$t$} & p-value & \multicolumn{1}{c}{$t$} & p-value & \multicolumn{1}{c}{$t$}\\
		 \midrule
			Disney 	&	 \textbf{0.004} 	& 3.0781 & \textbf{0.0389} 	& 2.1457 	& 0.151 		& 1.468  \\
			Marvel 	&	 0.1047 			& 1.6669 & 0.0588 			& 1.9554 	& 0.7195 		& 0.3621 \\
			Funny 	&	 \textbf{0.0001} 	& 4.3845 & \textbf{0.0042} 	& -3.0751 	& \textbf{0.0} & 8.6225 \\
			Strong 	&	 \textbf{0.011} 	& 2.6935 & 0.1751 			& -1.3859 	& \textbf{0.0} & 6.4942 \\
		\bottomrule
	\end{tabular}
	 \caption{$t$ test results for precision, recall, and F1 score. 
	 Out of 12 findings, 7 are statistically significant. In all cases, df = 33.}
	\label{tab:ttestaccuracy}
\end{figure}

\looseness-1 On average, we found \squid to be more effective in generating
accurate results than \sql (Figure~\ref{fig:accuracy}). For all four tasks, on
average across participants, results obtained with \squid achieved
significantly higher precision than the results obtained with \sql. \squid
achieved higher recall than \sql for the two objective tasks (Disney and
Marvel). While \squid's recall for the subjective tasks (Funny and Strong) was
lower than \sql, note that \sql's precision for those tasks was close to 0.
This is simply because the \sql queries the participants wrote for those tasks
were very imprecise and returned a very large number of results (e.g., all
actors in the database). While such general queries can happen to contain a
large portion of the correct results (hence the high recall), they contain an
extremely large number of irrelevant results making them poorly suited for this
retrieval task. In terms of F1 score, \squid always achieved higher values than
\sql implying its effectiveness over \sql for generating more accurate results.
The result of t-tests for these findings are shown in
Figure~\ref{tab:ttestaccuracy}. Out of the 12 findings, 7 are statistically
significant with a p-value less than 0.05.

\subsection*{Participants were more efficient with \squid than \sql}
\begin{figure}[t!]
	\centering
	\begin{subfigure}{.41\textwidth}
		\centering
		\includegraphics[width=1\linewidth]{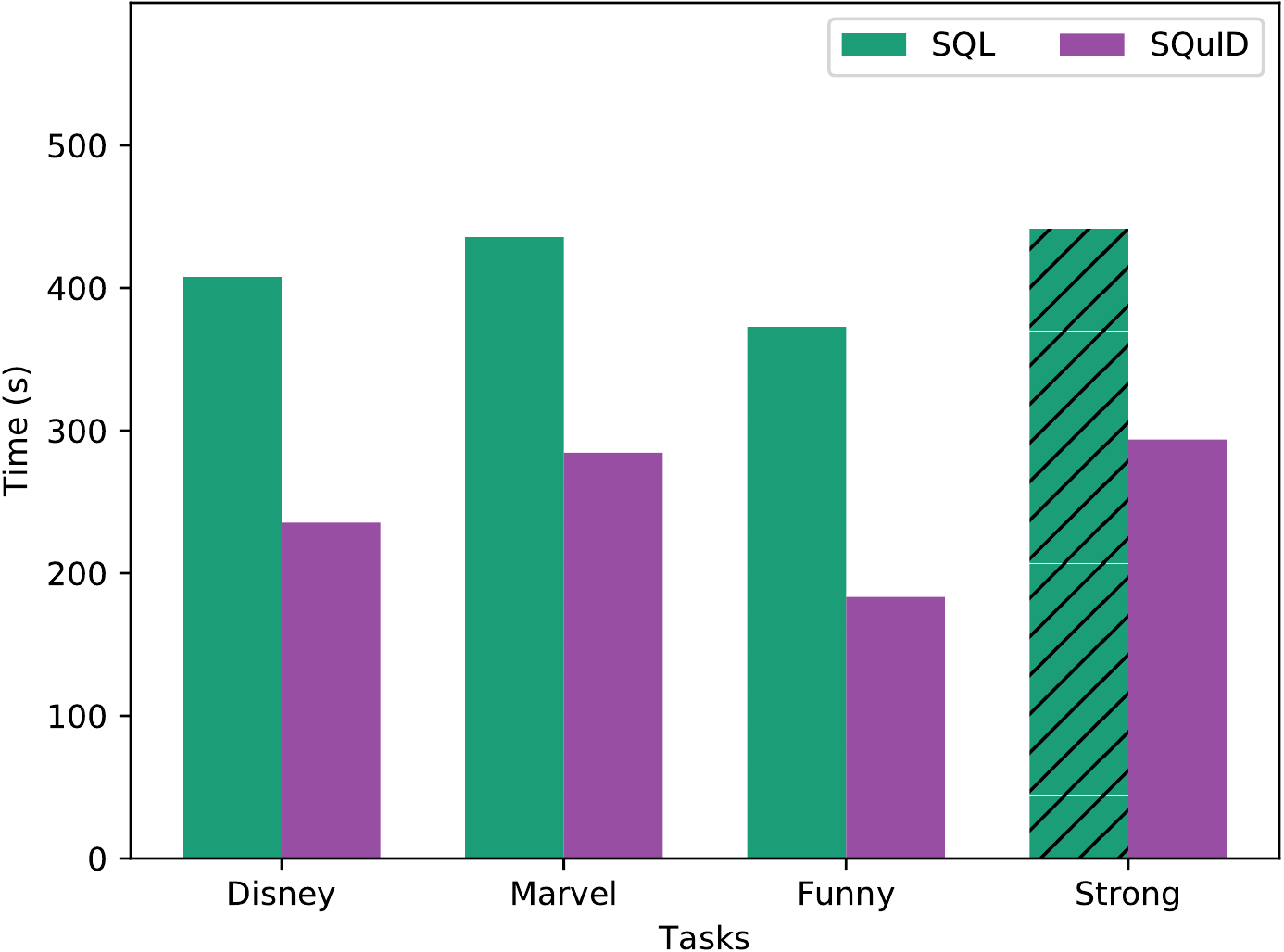}		
		\caption{Time required to solve each task.}
		%\vspace*{4mm}	 
		\label{fig:time}
	\end{subfigure}
	\hspace{4mm}
	\begin{subfigure}{.41\textwidth}
		\centering
		\includegraphics[width=0.98\linewidth]{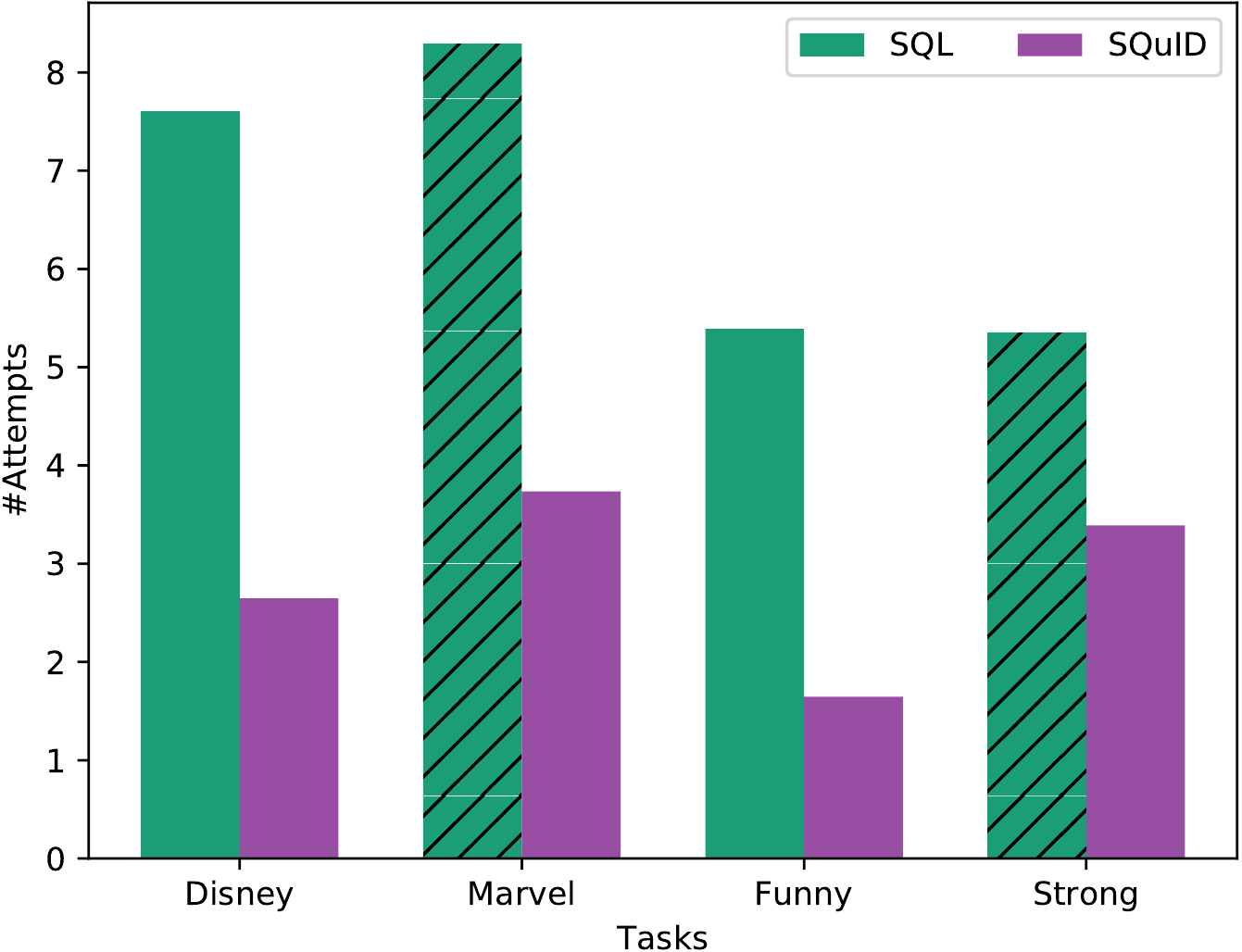}		
		 \caption{Number of attempts required to solve each task.}	 
		 %\vspace*{4mm}	 
		\label{fig:attempt}
	\end{subfigure}
	\caption{Comparison of \sql vs. \squid in terms of effort (average time
required and average number of attempts) for solving the same set of tasks.}
	\centering
	\begin{tabular}{l|ll|ll}
		\toprule
		\multicolumn{1}{c}{} & \multicolumn{2}{c}{Task completion time} & \multicolumn{2}{c}{\#Attempts} \\
		\midrule
		 \multicolumn{1}{c}{Task} & p-value & \multicolumn{1}{c}{$t$} & p-value & \multicolumn{1}{c}{$t$} \\
		 \midrule
			Disney & \textbf{0.0014} & -3.5000 & \textbf{0.0} & -4.7578 					 \\
			Marvel & \textbf{0.0146} & -2.5767 & \textbf{0.0008} & -3.6985 			   \\
			Funny & \textbf{0.0008} & -3.7105 & \textbf{0.0007} & -3.7441 				\\
			Strong & \textbf{0.0132} & -2.6206 & 0.0595 & -1.9518 						\\
		\bottomrule
	\end{tabular}
	\vspace{-2mm}
	\caption{$t$ test results for task completion time and number of attempts.
	Out of the 8 findings, 7 are statistically significant. In all cases, df = 33.}
	\vspace{3mm}
	\label{tab:ttesttime}	
\end{figure}

\looseness-1 \squid helped the participants solve the tasks more quickly
(Figure~\ref{fig:time}) and with fewer attempts (Figure~\ref{fig:attempt}) than
\sql. On average, the participants were able to solve the tasks using \squid
about~200 seconds faster than when using \sql. Participants were also able to
solve the tasks with about~4 fewer attempts while using \squid compared to
\sql. The results of t-test of these findings, shown in
Figure~\ref{tab:ttesttime}, signify that most are statistically significant
with a p-value less than 0.05.

\subsection*{Participants generally found \squid easier to use and more
satisfying, but still preferred \sql}

\begin{figure}[t]	
	\begin{subfigure}{.48\textwidth}
		\centering
		\hspace*{\fill}		
		\includegraphics[width=1\linewidth]{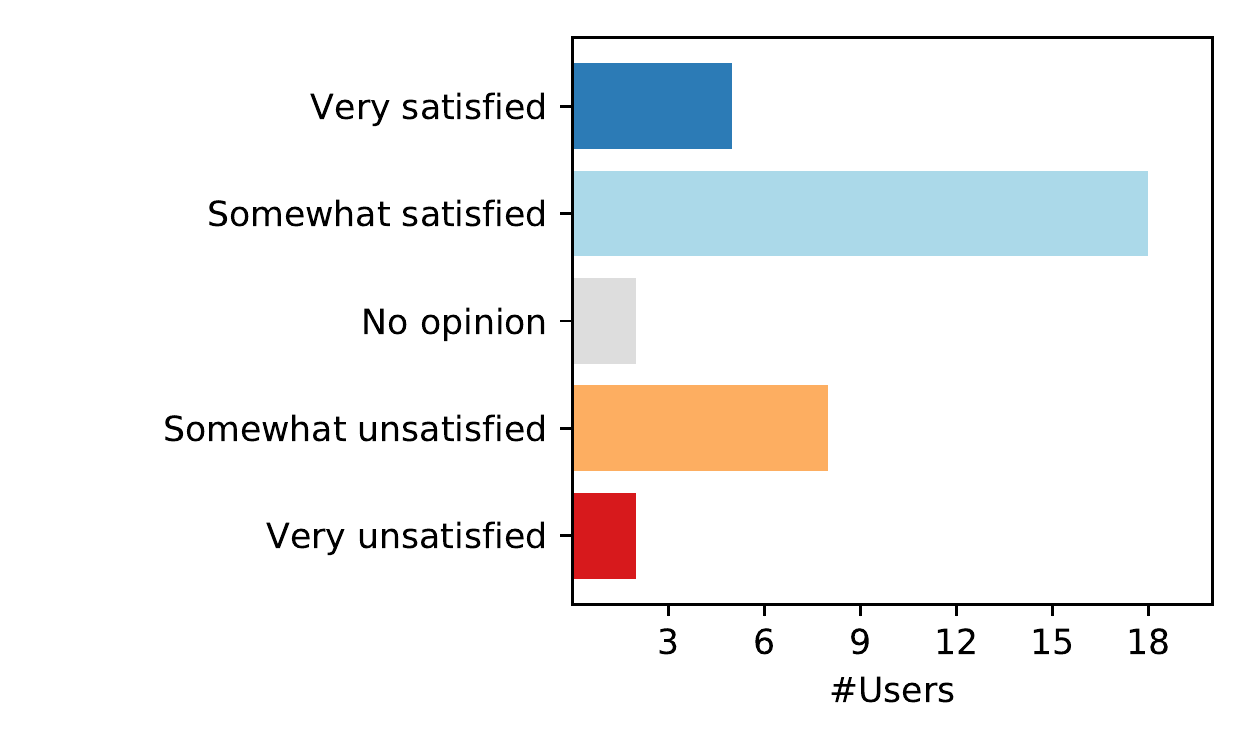}		
		 \vspace{-6mm}
		\caption{Overall satisfaction with \squid results.}	 
		\label{fig:squidsatisfaction}
	\end{subfigure}
	\hspace{3mm}
	\begin{subfigure}{.48\textwidth}
		\centering
		\hspace*{\fill}
		\includegraphics[width=1\linewidth]{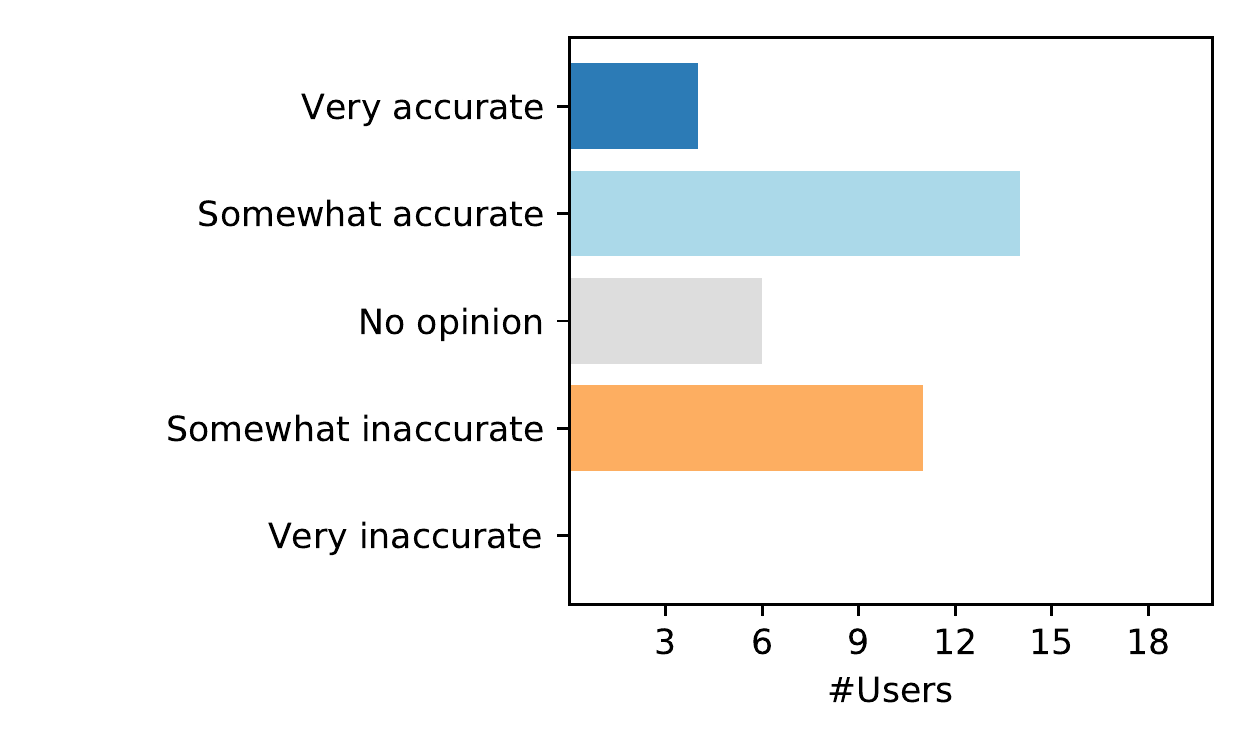}		
		 \vspace{-6mm}
		\caption{Overall accuracy of \sql results.}	 
		\label{fig:sqlsatisfaction}
	\end{subfigure}
	\begin{subfigure}{.48\textwidth}
		\centering
		\hspace*{\fill}
		\includegraphics[width=1\linewidth]{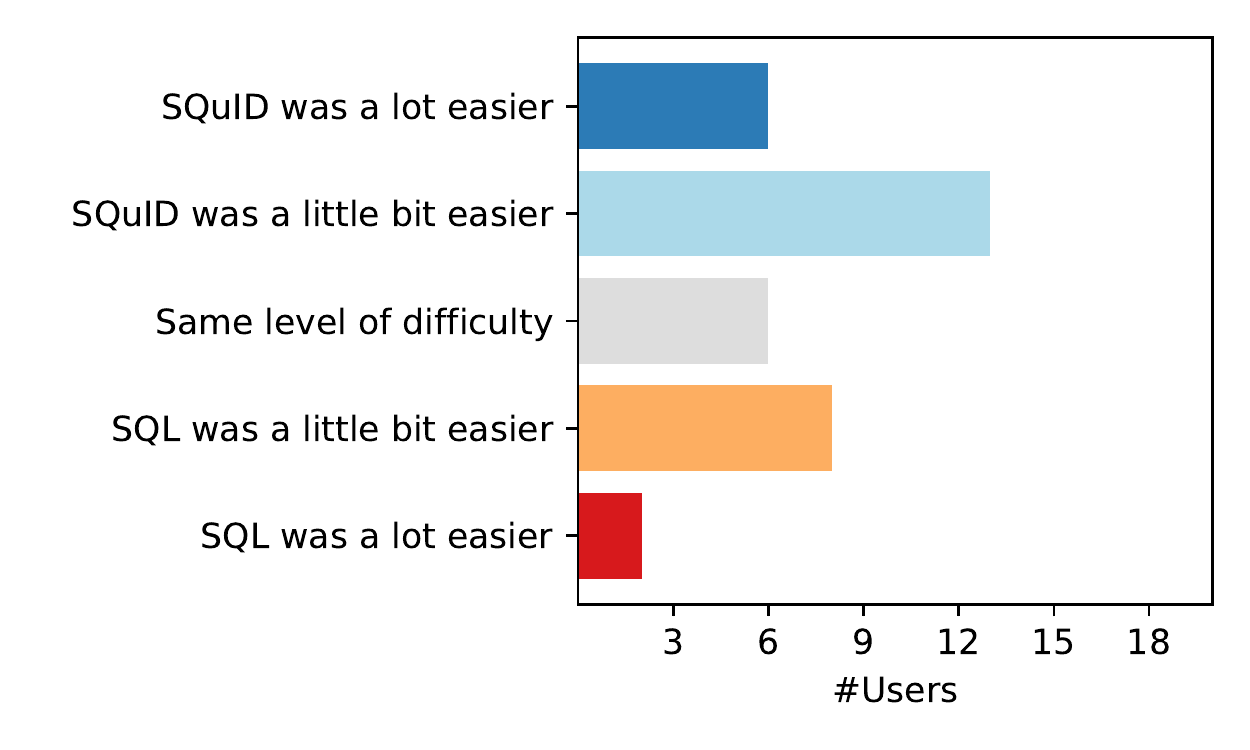}	
		 \vspace{-8mm}	
		 \caption{Usability}	
		 \vspace{-2mm}
	\label{fig:ease}
	\end{subfigure}
	\hspace{3mm}
		\centering
		\hspace*{\fill}
		\begin{subfigure}{.48\textwidth}
			\includegraphics[width=1\linewidth]{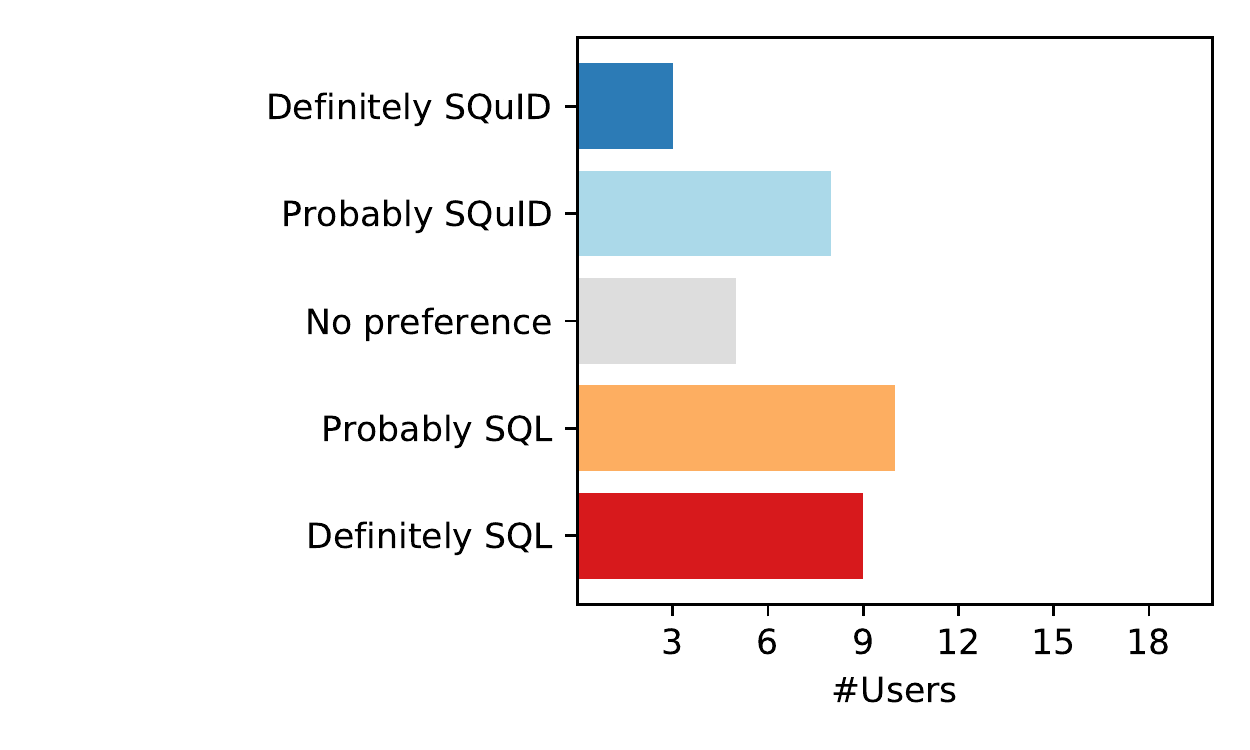}		
		 \vspace{-8mm}
		 \caption{Preference} 
		 \vspace{-2mm}
	\label{fig:preference}
	\end{subfigure}
	\caption{Comparison of \squid vs. \sql in terms of satisfaction with the results, overall ease of use, and preference (self-reported).}
	\vspace{-3mm}
\end{figure}

Figures~\ref{fig:squidsatisfaction} and~\ref{fig:sqlsatisfaction} show
self-reported overall satisfaction with the results produced by \squid and
\sql, respectively. Generally, participants found the results produced by \squid more
satisfying than the results produced by \sql. Out of the 35 participants, 23
were somewhat or very satisfied with \squid. In contrast, 18 reported that the
results produced by \sql were somewhat or very accurate. However, we found that
the self-reported satisfaction does not correlate with the actual correctness
of the results (measured in terms of precision, recall, and F1 score), and in
fact, the participants generally did better with \squid than \sql, although
they did not always realize it. Figure~\ref{fig:ease} shows self-reported
overall evaluation comparing \squid and \sql in terms of ease of use. Out of
the 35 participants, 19 reported that \squid was easier, 6 reported that they
had the same level of difficulty, and 10 reported that \sql was easier.

However, despite reporting that \squid was easier to use and the results were
more satisfying, the participants were still leaning towards \sql as a
preferred mechanism for data exploration. Figure~\ref{fig:preference} shows
self-reported overall preference between \squid and \sql, where 11 reported
that they would prefer \squid while 19 reported that they would prefer \sql.
Five participants reported no preference.

\section{Qualitative Feedback From Interview Study}\label{sec:qual}
We now report the results of our interview study and describe six main themes
that emerged from our qualitative analysis.

\subsection*{Studying the schema is challenging, even for \sql experts} All
seven of our interviewees from the interview study commented that it was
difficult to become acquainted with the database schema. ``As a user, I have to
\emph{explore} the schema'', I1 said. I1 continued, ``The query itself was not
complicated. It was time consuming to get familiar with the schema itself. Even
for experienced users, reading through the schema and getting acquainted to
[it] \dots takes time.'' When asked about the comparison in difficulty between
writing the \sql query and understanding the schema, I3 said ``Looking at the
schema diagram was harder. I kept going back and forth trying to understand
it.'' Understanding the schema may be complicated not only because it can be
difficult to learn what keys connect the tables, but also because it may be
hard to interpret the structure of the individual tables. I5 said, ``I think it
was pretty hard because I was not sure where to look for comedy based on
actors. I was thinking that [the] Role [table] might have the attribute, but it
didn't. Then I had to go through joining five tables!''

\subsection*{\sql requires stricter syntax, which makes writing queries hard}
\looseness-1 All interviewees struggled to a varying degree to write a \sql
query because of different issues; e.g., some of them could not figure out the
correct spelling of attributes. For instance, one would query for the genres
`scifi' or `comedic', neither of which exist in the database. I4 said, ``The
difficult part was to get the accurate predicate for the query, and I had to
[explore the database] for that.'' \sql requires strict string matching, which
can be extremely difficult to overcome for someone who is unfamiliar with the
database constants and \sql syntax. While it is possible to query a table and
view its content to see how the names are spelled, very few interviewees did
this. It appears that the ability to write a \sql query is based on experience
and recent exposure to \sql. Interviewees noted that they do not use \sql on a
daily basis---some even said they had not used \sql in months---thus, it was
difficult to recall specific syntax. For instance, two of the
interviewees---who had relatively lower \sql expertise---could not remember the
requirements for joining tables. I7 had to use Google to help with this syntax,
and I2 did not recall that \sql could join more than two tables. I5 said, ``I
was making a lot of mistakes about where to have the underscores, where to not
have underscores, and for those things I had to look through the [schema]
multiple times.'' An interesting note, I6 spent the vast majority (over 9
minutes) attempting to find the name `Japan' in the database, and spent less
than 1 minute writing the actual query. \squid reduces the need to recall exact
spelling by providing an auto-completion feature as the user types examples.
Although it does not provide an auto-correct if the name is spelled
incorrectly, the auto-completion feature allows the users to type what they
know and scroll through the suggestions until they find the proper name. We
observed several of our interviewees initially spelled a movie name
incorrectly, but they were helped by the auto-completion feature. For example,
I2 initially typed `Spiderman' in the search bar, but the title is spelled
`Spider-Man' in the database. I2 was able to correct the spelling when he typed
`Spider' in the search box and autocomplete showed the entire title. The search
bar also helped I5 who noted, ``If I was missing some spellings, there were
some suggestions.''

\subsection*{\sql requires parameter tuning for subjective tasks; \squid alleviates this}
Some exploration tasks can be subjective and inherently vague, e.g., how does
one define a ``funny'' actor \emph{precisely}? How many comedies, exactly, does
an actor have to star in before they are considered funny? These questions have
no clear answers, and such parameters can vary from person to person and from
day to day. In practice, it may be very difficult, if not impossible, to think
of objective measures for a subjective concept, which makes subjective tasks
very complicated to specify with \sql. I2 said, ``Even if I forget about syntax
\dots figuring out how to go about writing the pseudo code query for funny
actors [is difficult]''. One of the most common blunders of interviewees who
used \sql to find ``funny'' actors was to query all actors who had been in some
comedy movie. I3 was the first to acknowledge this. ``I had to play around with
a lot of smaller queries,'' he said, ``to get the one that I eventually had,
which I was still not satisfied with. It seems like I pulled many actors and
actresses that happened to be in some comedy.'' I3 elaborated, ``Vague tasks
are generally a lot more open to interpretation. Coding up a query that meets
someone's vague specifications [is] hard \dots It was very hard to nail down
what the correct definition of funny is.'' I4 also recognized that vague tasks
are difficult to define. She even said, ``This probably isn't a query that I
should write in \sql!'' She continued, ``strong and muscular are very vague
descriptors, and \sql needs clear rules. I have to use genre as a proxy, and
that makes the query very nasty.''

On the other hand, \squid can interpret complex parameters without any
involvement from the user, sparing them the mental burden of defining and
implementing a complex query. I4 also said, ``In order to write a \sql query,
you need to understand the schema well, know your data well, and know your
question well \dots But if the task is exploratory and you only have a vague
idea in mind, like `strong actors' \dots it would be very hard, if not
impossible, to write a \sql query.'' Indicating how \squid helped in the
subjective tasks, I3 said ``\squid is a lot more user-oriented. You could just
put in some actor names and it would infer what you really want.''

\subsection*{\squid produces precise results, which is preferred for data
exploration} We asked interviewees whether they would prefer a long list that
includes all relevant names, but may also include many irrelevant names
(high-recall) or a shorter list that includes exclusively relevant names with
very few irrelevant names, but may miss some relevant names (high-precision).
Six out of seven interviewees reported that they would prefer having a shorter
list with higher precision, while one interviewee had no preference. ``I think
I'm okay with not having all Marvel movies listed here,'' I2 said, ``but I
definitely don't want anything outside of Marvel movies. It's fine that [the
results] are missing some Thor movies. I wouldn't have liked it if there were
movies from DC [Comics] in here.'' Comparing the \sql results to the \squid
results, I5 said, ``I think the [\squid] results were not too few but not too
many. It was easily understandable, and I could actually see if these were
actors I was looking for \dots The [\sql] results were just too many, and most
of the names I didn't know, so it was not easy to find the names that I was
looking for.'' I6 said, ``I prefer a shorter list because if there are too many
movies listed, then probably, it would be overwhelming and I could not say if
the results are right.''

\subsection*{\squid's interactivity aids users to enrich examples} Three
interviewees mentioned that the results produced by \squid helped them think of
more examples in an iterative process. I6, who struggled to think of examples,
was able to think of only three sci-fi movies, but when she saw `Avatar' in the
list of results, many other ideas came to her mind. Even if the intermediate
results (the first or second round of results generated) were not all intended,
some of them were useful in reminding the interviewees of relevant examples.
For instance, I2 said, ``\squid was [nice] because it was slightly interactive.
I could look at the results and update my examples.'' During a task, I7 said,
``[The results are] useful because now I can use Guardians of the Galaxy.'' I7
later added, ``I think when I gave the first few examples, it gave me some
results and that helped me think of more that I was looking for, and it
eventually did complete the task.'' \squid's results reminded the interviewees
of examples that hadn't been in the forefront of their mind, but were
nonetheless relevant. I3 said, ``I saw the movie Transformers, and that's
something I had in my mind, but it did not occur to me when I was entering the
examples. There were a bunch of other movie names [like that].'' Since \squid
can provide serendipitous, but helpful, intermediate results, the user's lack
of domain familiarity can still be alleviated to some extent.

\subsection*{Domain familiarity is crucial to evaluate the results, for both
\squid and \sql } \squid requires a basic familiarity with the domain. For
those who struggle to think of even one relevant example, like I6, \squid
presents a unique challenge. All interviewees could easily think of a few
examples that fit the task, but they struggled beyond that. I7 said, ``It was
very easy to come up with two or three, but the more examples I had to give the
harder it became''. Two interviewees suggested that \squid adopt an interactive
system where it would ask the user whether or not a particular result was
relevant on a case-by-case basis. This could alleviate some of the difficulty
of thinking of relevant examples.

Furthermore, users who possess very little knowledge of the domain may be
unable to recognize the results, and thus would be incapable of verifying them.
But this is true for both \squid and \sql. It was not uncommon for the
interviewees to tell us that they could hardly recognize the names in the
results, especially for \sql. I1, for instance, said, ``Honestly, I don't
recognize any of the results.'' This, apparently, was partly due to the large
number of results returned by \sql, where there is a high chance that there will be
unfamiliar names. Most people are only familiar with a relatively small subset
of actors, rather than the entirety of the IMDb database. This made it
difficult for the users to evaluate the results produced by both \squid and
\sql.

\section{Discussion and Future Work}\label{sec:discussion}
In this section, we summarize significant findings found from the quantitative
and qualitative analysis of our comparative user studies and highlight the key
take-aways.

\subsection*{\squid alleviates \sql pain-points: schema complexity, semantic
translation, and syntax}

From our interviews, we identified three key pain-points of the traditional
\sql querying mechanism, all of which are removed when using \squid:

\subsubsection*{Schema complexity} One significant difficulty that we observed
during the use of \sql was the requirement of schema understanding. To issue a
\sql query over a relational database, the user must first familiarize
themselves with the database schema~\cite{DBLP:journals/pvldb/FarihaM19,
ShenSIGMOD2014, DBLP:conf/sigmod/BaikJCJ20}. The schema is often complex, such
as the IMDb schema shown in Figure~\ref{fig:schema}, and requires significant
effort to understand. The user also needs to correctly specify the constant
values (e.g., \texttt{Comedy} and not \texttt{Comedic}), name of the relations
(e.g., \texttt{movietogenre} and not \texttt{movie\_to\_genre}), and name of
the attributes (e.g., \texttt{id} and not \texttt{movie\_id}) in the \sql
query. Moreover, some attributes reside in the main relation (e.g.,
\texttt{person.name}) while others reside in a different relation (e.g., names
of a movie's genres reside in the relation \texttt{genre} and not in the
relation \texttt{movie}). From a closer look at some of the user-issued \sql
queries, we observed futile efforts to guess keywords, incorrectly trying
values such as ``comedic'', ``superhero comics'', and ``funny'', which do not
exist in the database and result in syntax or semantic errors. In structured
databases, if one does not know the exact keywords, they end up issuing an
incorrect \sql query, which returns an empty result. In contrast, \squid frees
the user from this additional overhead as it leverages the database content and
schema and associates it automatically with the user-provided examples.

\subsubsection*{Semantic translation} \looseness-1 After studying the schema,
the next task was to translate the task's semantics formally to a language
(e.g., \sql) that computational systems understand. While this is relatively
easy for objective tasks (e.g., finding all movies produced by Disney), the
same is not true for subjective tasks (e.g., finding all ``funny'' actors). As
our qualitative feedback indicates, expressing subjective or vague tasks is
hard in any formal language and not only in \sql. For example, for the task of
finding all ``funny'' actors, even the \sql experts struggled to encode the
concept ``funny'' in \sql. Many participants wrote a \sql query to retrieve all
actors who appeared in at least one movie whose genre is \texttt{Comedy}.
However, upon observing the output of such an ill-formed query, they were not
satisfied with the results. This is because appearing in only one comedy movie
does not necessarily make an actor funny. Usually, actors who appear in
``many'' comedy movies are considered funny. The key struggle here is to figure
out what is the right threshold for ``many'', i.e., in \emph{how many} comedy
movies should an actor appear to be considered ``funny''. In contrast, \squid
is able to discover these implicit constants from the user-provided examples.
For retrieving funny actors, \squid learns from the user-provided examples what
is the usual number of comedy movies all the example actors appeared in, and
subsequently, uses that number to define the notion of ``many''. For instance,
in the usage scenario of Section~\ref{squid}, \squid inferred that appearing in
$40$ comedy movies is sufficient for an actor to be considered funny. This
parameter ($40$) was automatically inferred based on the user-provided
examples: \squid automatically discovered that each example actor appeared in
$40$ or more comedy movies in the IMDb database.

\subsubsection*{Language syntax}
\sql is a programming language with several operators and keywords, and
similar to all programming languages, \sql also requires strict syntax. While
issuing a \sql query, even a minor syntactic error will result in complete
failure and will return no result. Moreover, the syntax error messages that the
\sql engine provides are often ambiguous and confusing to novice users. We
observed that one of our interviewees could not recall the correct syntax of
the \texttt{JOIN} operation. This stringent requirement of syntax poses significant
hurdles to novice and even intermediate \sql users. In contrast, \squid
completely bypasses \sql, eliminating this challenge.

\subsection*{\squid is generally more effective than \sql and boosts efficiency}
In our controlled experiments, we noted that \squid is generally more effective
than \sql in deriving accurate results. For objective tasks, we found that
\squid outperforms \sql in all three correctness metrics---precision, recall,
and F1 score. However, it is important to highlight that our interviewees noted
that \squid is particularly useful and preferable to \sql for \emph{subjective}
tasks. This does not contradict our quantitative analysis. While \sql has
higher recall than \squid for subjective tasks, \squid achieves much higher F1
scores, because \sql's precision for these tasks is close to 0. This is because
an extremely general \sql query (e.g., one that returns all the data) may have
very high recall, but it will not be of use to the exploration task that
expects targeted results. Furthermore, \squid significantly boosts the user's
efficiency in data exploration. This was confirmed by our controlled experiment
study where we found that participants achieved their goal much faster (in
about 200 fewer seconds) and with less effort (with about 4 fewer attempts)
while using \squid compared to \sql.

Lack of domain knowledge is a handicap for \squid, as it requires at least a
few initial examples for its inference. This is a general issue with all
query-by-example mechanisms~\cite{DBLP:journals/pvldb/FarihaM19,
ShenSIGMOD2014, DBLP:conf/popl/Gulwani11}. However, even when the user lacks
domain knowledge, they can use alternative mechanisms---such as keyword search,
Internet search, or very basic \sql queries (when the user has some \sql
familiarity)---to come up with some initial examples. In contrast, when a user
does not know \sql, learning it from scratch takes significant time and effort.
While \squid's by-example paradigm can help both expert and novice users alike,
in general, programming-by-example systems are most beneficial when domain
knowledge outweighs technical knowledge and
experience~\cite{DBLP:conf/oopsla/SantolucitoGWP18}; otherwise, a hybrid system
is more desirable. However, lack of domain knowledge is a problem for \sql as
well. Without basic knowledge over the data domain (e.g., what are the entities
and what are their properties), understanding the schema can be harder.
Furthermore, without sufficient domain knowledge, debugging \sql queries, i.e.,
validating whether the user-issued \sql queries are correct or not, based on
the results, is also challenging.

\subsection*{\squid promotes serendipitous discovery, aiding users in data
exploration} \looseness-1 \squid is \emph{interactive} in a sense that the
users can revise their examples based on the results and even use some of the
results as examples in the next iteration. A number of interviewees mentioned
that by looking at the results that \squid generated from their initial
examples, they were able to come up with new examples. Moreover, when their
examples contained some unintentional bias---e.g., while retrieving Disney
movies, they only provided examples of recent movies---they were able to
receive implicit feedback of that bias by \squid as the results \squid
generated reflected the same bias. This feedback mechanism helped them revise
their examples accordingly. In contrast, \sql does not offer such interactivity
or feedback mechanism. While some interviewees used subqueries of the main
query to view some intermediate results, this was just for the purpose of
verifying the correctness of the main query. In contrast, \squid's natural
interaction and feedback mechanism offers additional help to the users. This
makes \squid particularly suitable for the task of data exploration. \squid
often promotes \emph{serendipity} in the results---providing a good balance
between \emph{exploration} (serendipitous, surprising, and novel discovery) and
\emph{exploitation} (similar to the examples)---which is a desired property
during data exploration.

\subsection*{\squid is particularly useful for solving complex and subjective
tasks} \looseness-1 The specific properties of \squid, specifically
interactivity, providing feedback, and promoting serendipitous discovery, make
it a significantly better choice for solving subjective tasks that are usually
ambiguous and vague, and are very hard to solve using \sql. For example, in our
studies, we used ``strong actors'' or ``funny actors'' as two examples of
subjective tasks. Participants of both our controlled experiment study and
interview study found thinking of examples easier than expressing their intent
using \sql, especially for subjective tasks. Our results indicate that \squid
provides an easier mechanism for data retrieval and helps users overcome the
difficulty of writing overly complex \sql queries for subjective tasks. In
contrast, for objective tasks, we found both \squid and \sql equally effective,
given the user has basic \sql expertise.

\subsection*{Trust on a system depends on prior exposure, expertise, type of
the tasks, and system explainability}

During our controlled experiment, we wanted to measure how much the
participants trust the mechanism that produces the results by asking the
questions: ``how well do you think \squid did in generating the desired
results?'' and ``how accurate were the \sql results?'' While some participants
reported that they were more satisfied with the results produced by \squid than
\sql, interestingly, many of them reported that they prefer \sql over \squid
even though they generally did better with \squid
(Figure~\ref{fig:preference}). This result is in line with prior work that
compared a PBE tool against traditional shell-scripting and found that despite
performing better using the PBE tool, users tend to trust the traditional
shell-scripting more~\cite{DBLP:conf/oopsla/SantolucitoGWP18}. We validated
this by checking against ground-truth results where \squid groups achieved
results with higher precision (more specific) and F1 score (more accurate), as
shown in Figure~\ref{fig:accuracy}.

\looseness-1 Since the participants performed better when using \squid compared
to \sql, we interpret their preference for \sql to be due to three possible
sources of bias: (1)~\emph{Familiarity:} The participants were at the time
taking a course on relational databases and \sql, which may have artificially
increased their confidence in their \sql skills. They had prior experience with
\sql, but were experiencing \squid for the first time through the study.
(2)~\emph{Explainability:} \sql exposes the precise mechanism (the code) that
produces the results, while we did not provide participants with an explanation
of the inner workings of \squid nor exposed the query it produces.
(3)~\emph{Domain expertise:} Low domain expertise poses a hurdle in producing
examples for \squid; we posit that the users may consider \sql a more versatile
mechanism for such circumstances.

We further investigated the issue of trust during our interview study by asking
all our interviewees the question: ``Which of these two systems, \squid or
\sql, do you trust more?'' We expected \sql experts to trust \sql more, but did
not observe any strong trend. Rather, the interviewees mentioned that for
objective tasks, they were more confident about the \sql queries they wrote,
and hence, they trusted \sql more. In contrast, for the subjective tasks, they
reported that they trusted the results produced by \squid more, as for the
subjective tasks, the most common complaint was that \sql produced too many
results (less specific) and perhaps retrieved the entire database content.
Ultimately, \squid can also provide explanations, by exposing the \sql query it
synthesizes in order to generate the results and the underlying mechanism used
to synthesize the query. We shed more light on this in the future work.

\subsection*{\squid is easy to learn} A desired property for any system is
\emph{learnability}: how easy it is to get used to the system. From our study,
we found that it was very easy for the participants to learn how to use \squid
almost instantly. \squid's interface is intuitive and both novices and experts
learned how to use it, just by observing its behavior. In contrast, when
participants did not know how to write certain classes of \sql queries, they
simply gave up and mentioned that they cannot express their logic in \sql. This
is particularly significant considering that all our study participants and
interviewees had prior exposure to and experience with \sql, while this was
their first experience using \squid.

\subsection*{Limitations and future work}
Our study results indicate that \squid effectively helped users with various
levels of \sql familiarity perform their tasks faster and more efficiently.
However, our work explored only one example of QBE systems and recognizably
with a limited number of participants. Additional work is needed to study the
impact of QBE systems further. While our goal was to draw a comparison between
traditional and QBE systems, additional studies might investigate how complete
novices (users with no \sql expertise) use QBE systems. Furthermore, future
studies can expand the list of tasks to tease apart better the impact of using
QBE systems for various task types. From the interviewees' feedback, we
extracted a few directions for future work to improve user experience while
using QBE systems:

\subsubsection*{Exposing the synthesized \sql query for explainability} One
shortcoming of \squid is that the user is unaware of the mechanism \squid uses
to generate the results. Under the hood, \squid synthesizes a \sql query from
the user-provided examples, which it uses to produce the results. A possible
future work for QBE systems is to expose the \sql query and allow the users to
fine tune the query parameters to suit their specific purposes.

\subsubsection*{Exposing internal mechanism for further explainability} In
addition to exposing the \sql query, QBE systems can provide further
explanation mechanisms by exposing the particular semantic similarities that
the system discovers across the examples, and its confidence in each similarity
being intended. This can also guide users in revising their examples to
emphasize borderline semantic similarities that \squid missed, or diversify
examples to avoid coincidental similarities among the examples.

\subsubsection*{Tuple suggestion to enrich examples} A few interviewees
reported that it would be helpful if \squid could suggest a few tuples that the
user may consider adding to the examples. Such a tuple-suggestion mechanism
will help the users supply additional examples and diversify the examples, in
case the users lack domain knowledge.

\subsubsection*{Interaction with the results for feedback} Another direction of
future work is to allow the users to interact with the results produced by QBE
system: the user will accept or reject a few result tuples which will act as
feedback to the system. This will help QBE system learn the user intent better.

\subsubsection*{Extensive user study} More extensive user studies are needed in
the future to evaluate all these additional features and determine whether they
contribute positively to the users' trust and satisfaction in QBE systems.

\section{Conclusions}\label{sec:summary}
Our comparative user studies found that database users, with varied levels of
prior \sql expertise, are significantly more effective and efficient at a
variety of data exploration tasks with \squid over the traditional \sql
querying mechanism that requires database schema understanding and manual
programming. Our results indicate that \squid eliminates the barriers of
familiarizing oneself with the database schema, formally expressing the
semantics of an intended task, and writing syntactically correct \sql queries.
The key take-away of this work is that in a programming-by-example tool like
\squid, even a limited level of domain expertise (knowledge of a subset of the
desired data) can substantially help overcome the lack of technical expertise
(knowledge of \sql and schema) in data exploration and retrieval. This
indicates that programming by example can lead to the democratization of
complex computational systems and make these systems accessible to novice users
while aiding expert users as well. Our studies validate some prior results over
other PBE approaches but also contribute new empirical insights and suggest
future directions for QBE systems to further increase system explainability and
user trust.

\bibliographystyle{ACM-Reference-Format}
\bibliography{paper}

\end{document}